\documentclass[12pt,nofootinbib,tightenlines,preprintnumbers,superscriptaddress]{revtex4}
\usepackage[T1]{fontenc}
\usepackage[utf8]{inputenc}
\usepackage{lmodern}
\usepackage[english]{babel}
\usepackage{graphicx}
\usepackage{amsmath,mathtools,amssymb}
\usepackage{hyperref}
\usepackage{tabularx}
\usepackage{hhline}
\usepackage{booktabs}
\usepackage[dvipsnames]{xcolor}
\usepackage{tikz}
\usepackage{bm}
\usetikzlibrary{decorations.pathreplacing,decorations.pathmorphing}

\usepackage{siunitx}
\usepackage{soul}

\newcommand{\sd}{\mathrm d}

\newcommand{\spert}{\mathrm {pert}}
\newcommand{\sfit}{\mathrm {fit}}
\newcommand{\Cov}{\mathrm {Cov}}
\newcommand{\beq}{\begin{equation}}
\newcommand{\eeq}{\end{equation}}
\newcommand{\bit}{\begin{itemize}}
\newcommand{\eit}{\end{itemize}}
\newcommand{\ben}{\begin{enumerate}}
\newcommand{\een}{\end{enumerate}}

\usepackage{physics}

\newcommand{\om}{\omega}

\DeclareFontFamily{U}{mathx}{\hyphenchar\font45}
\DeclareFontShape{U}{mathx}{m}{n}{<-> mathx10}{}
\DeclareSymbolFont{mathx}{U}{mathx}{m}{n}
\DeclareMathAccent{\widebar}{0}{mathx}{"73}

\preprint{MITP-22-032}
\preprint{CERN-TH-2022-070}

\begin{document}

\title{Photon emissivity of the quark-gluon plasma:\\ a lattice QCD analysis of the transverse channel}

\author{Marco C{\`e}}
\affiliation{Albert Einstein Center for Fundamental Physics (AEC)
and Institut für Theoretische Physik,
Universität Bern, Sidlerstrasse 5, CH-3012 Bern, Switzerland}
\affiliation{Theoretical Physics Department,
CERN,
CH-1211 Geneva 23, Switzerland}

\author{Tim Harris} 
\affiliation{School of Physics and Astronomy,\\
University of Edinburgh, EH9 3JZ, UK}

\author{Ardit Krasniqi} 
\affiliation{PRISMA$^+$ Cluster of Excellence \& Institut f\"ur Kernphysik,
Johannes Gutenberg-Universit\"at Mainz,
D-55099 Mainz, Germany}

\author{Harvey~B.~\!Meyer} 
\affiliation{PRISMA$^+$ Cluster of Excellence \& Institut f\"ur Kernphysik,
Johannes Gutenberg-Universit\"at Mainz,
D-55099 Mainz, Germany}
\affiliation{Helmholtz~Institut~Mainz,
Johannes Gutenberg-Universit\"at Mainz,
D-55099 Mainz, Germany}
\affiliation{GSI Helmholtzzentrum f\"ur Schwerionenforschung, Darmstadt, Germany}

\author{Csaba T\"or\"ok} 
\affiliation{PRISMA$^+$ Cluster of Excellence \& Institut f\"ur Kernphysik,
Johannes Gutenberg-Universit\"at Mainz,
D-55099 Mainz, Germany}

\begin{abstract}
We present results for the thermal photon emissivity of the quark-gluon plasma derived from 
spatially transverse vector correlators computed in lattice QCD
at a temperature of 250\;MeV.
The analysis of the spectral functions, performed at fixed spatial momentum, is based
on continuum-extrapolated correlators obtained with two flavours of dynamical Wilson fermions.
We compare the next-to-leading order perturbative QCD correlators, as well as the ${\cal N}=4$
supersymmetric Yang-Mills correlators at infinite coupling, to the correlators from lattice QCD and find them
to lie within $\sim$10\% of each other. We then refine the comparison, performing it at the level
of filtered spectral functions obtained model-independently via the Backus-Gilbert method.
Motivated by these studies, for frequencies $\omega\lesssim2.5\,$GeV we use fit ans\"atze to the spectral functions
that perform well when applied to mock data generated from the NLO QCD 
or from the strongly-coupled SYM spectral functions, while the high-frequency part, $\omega\gtrsim 2.5\,$GeV,
is matched to NLO QCD. We compare our results for the photon emissivity
to our previous analysis of a different vector channel at the same temperature.
We obtain the most stringent constraint at photon momenta around $k\simeq0.8\,$GeV, for which we find
a differential photon emission rate per unit volume of 
$\sd\Gamma_\gamma/\sd^3k =   (\alpha_{\rm em}/(\exp(k/T)-1))\times (2.2 \pm 0.8 ) \times 10^{-3}\,{\rm GeV}$.
\end{abstract}

\maketitle

\section{Introduction}
\label{sec:intro}

Photons and lepton pairs have long been  considered
to provide direct information on the quark-gluon plasma (QGP), since they 
are penetrating probes of the QGP due to their colourless 
nature~\cite{Feinberg:1976ua,Shuryak:1978ij,Ferbel:1984ef,Cassing:1999es,David:2019wpt}.
During heavy-ion collisions, photons can escape the plasma without scattering via the strong 
interaction, but discriminating between different sources is 
quite challenging because of the continuous emission of photons during
the spacetime evolution of the fireball.
The detected photons are divided into two
main categories: decay and direct photons.
The former refers to photons coming from the electromagnetic decays of 
final state hadrons, while the latter includes all photons created in the 
collision before the final hadrons completely decouple.
The decay photons give a much larger contribution to the signal than the
direct photons and they provide valuable information for particle 
reconstruction. 
However, when studying direct photons, that contribution amounts to
a large background and has to be subtracted from the total photon yield.

Direct photons are produced via several mechanisms during the
evolution of a heavy-ion collision (c.f.~\cite{David:2019wpt}), but the two
major sources are initial hard parton-parton scattering (prompt photons),
and photons originating from the QGP (thermal photons).
The latest direct photon yield results have been published by the
PHENIX~\cite{Adare:2014fwh} and STAR~\cite{STAR:2016use} collaborations at RHIC,
and by ALICE~\cite{Adam:2015lda} at the LHC.
The direct photon yield measured by the PHENIX and by the ALICE
collaborations shows an excess at low transverse momentum,
$p_T \lesssim 3~\mathrm{GeV}$, with respect to 
the theoretical predictions~\cite{Paquet:2015lta,Gale:2021zlc}.
The results of the STAR collaboration~\cite{STAR:2016use}, however, are in 
better agreement with the model results.
The excess observed in the PHENIX and ALICE results is in the 
momentum range where the dominant contribution comes from thermal
photons~\cite{Gale:2021zlc}.

An important ingredient in the theoretical prediction is the emission
rate of thermal photons per unit volume, which has to be integrated over the expanding spacetime
volume of the medium to obtain the total thermal photon yield~\cite{Gale:2021zlc}.
This rate has been calculated at leading order in the strong coupling 
constant~\cite{Arnold:2001ms}, and the calculation has been extended to
include corrections which arise from interactions with soft
gluons~\cite{Ghiglieri:2013gia}.
The thermal photon emission rate from this extended calculation 
has a similar functional form to the leading order one, and 
represents a 20\% increase at $\alpha_\mathrm{s}\simeq0.3$.
This modest increase reduces the tensions, though it is still 
insufficient to explain the excess observed by the PHENIX and 
ALICE collaborations and highlights the leading-order prediction may receive
large corrections at relevant temperatures.

Therefore, a precise non-perturbative calculation of the thermal
photon production rate using lattice QCD is highly desirable
and would help to resolve the tensions or confirm the already existing 
weak-coupling results.
Although lattice QCD has been very successful in calculating 
observables in vacuum as well as in thermal equilibrium,
reliably accessing near-equilibrium quantities such as the thermal photon rate
is very challenging, because the calculation of real-time
observables requires analytic continuation using a finite, limited number
of noisy Euclidean correlator data.
Moreover, the Euclidean correlators that need to be inverted to
obtain information on the spectral function are rather insensitive
to the infrared features of these spectral functions~\cite{Aarts:2002cc,Teaney:2006nc,Meyer:2011gj,Borsanyi:2018srz}.
In spite of the difficulties, several methods have been devised
to constrain the ill-posed numerical inversion problem using both model-dependent and
independent approaches.

To determine the thermal photon production rate via analytic continuation,
the starting Euclidean observable is the vector current correlation function
at finite temperature.
Although this correlator has been investigated extensively also on thermal
ensembles, earlier analyses focused mainly on vanishing spatial momentum,
which is relevant for the determination of the electrical conductivity
of the plasma~\cite{Ding:2010ga,Brandt:2012jc,Amato:2013naa,Aarts:2014nba,Brandt:2015aqk,Ding:2016hua}.
Continuum-extrapolated vector current correlators
at finite spatial momenta obtained in quenched lattice-QCD simulations
were analyzed for the photon rate in Ref.~\cite{Ghiglieri:2016tvj}.
In Ref.~\cite{Ce:2020tmx}, the same goal was pursued
with continuum-extrapolated vector correlators based on
two-flavour dynamical simulations,
focusing on an infrared-dominated channel, the difference
between transverse and longitudinal channels ($\mathrm T-\mathrm L$ channel).
In both~\cite{Ghiglieri:2016tvj} and~\cite{Ce:2020tmx},
information on the spectral function was extracted from
the Euclidean data using fit ans\"atze.

In this work, we apply two different strategies to analyse the correlation
functions:
the Backus-Gilbert method~\cite{Backus:1968abc,Backus:1970} and a fit method where we apply various simple
fit ans\"atze matched to perturbation theory at high frequencies.
In contrast to our earlier work, we investigate the spatially transverse channel
of the vector current correlator at finite spatial momentum.
In Sec.~\ref{sec:theory}, we collect the relevant basic formulas and
discuss the advantages of investigating this particular channel.
We also present in that section an overview of the spectral function
in this channel in different regimes.
Details on the lattice configurations and the observables relevant in 
this study, including the continuum extrapolation are presented
in Sec.~\ref{sec:latres}.
The analysis of the continuum-extrapolated correlator in the transverse
channel is presented in Sec.~\ref{sec:estim}.
There, we also present our estimate on the thermal photon rate based on
this channel and compare it to available results from the literature.
We finally give our conclusions and an outlook in Sec.~\ref{sec:concl}.

\section{Theoretical background}
\label{sec:theory}

In this section, we collect the main definitions of the correlation
functions to be analyzed and describe the existing theory predictions
with which we will confront our lattice QCD calculations.

\subsection{Basic definitions}
\label{sec:basics}

The spectral function of the vector current $J_\mu= \bar\Psi \gamma_\mu T^a \Psi$,
for a generic matrix $T^a$ acting in quark-flavour space, is defined as
\beq
\rho_{\mu\nu}(\om, {\bf k}) = \int \sd t\,\sd^3 x \, \mathrm e^{\mathrm i(\om t - {\bf k \cdot x})}\,
\langle [J_{\mu}(t,{\bf x}), J_{\nu}(0)^\dag] \rangle.
\eeq
The Minkowskian time evolution of the current is given by
$J_\mu(t,{\bf x}) = \mathrm e^{\mathrm iHt} J_\mu(0,{\bf x}) \mathrm e^{-\mathrm iHt}$.
The~expectation value of the commutator is taken with respect to the thermal density matrix, $\mathrm e^{-\beta H}/Z$,
with $\beta = 1/T$ being the inverse temperature.
To leading order in the fine-structure constant, $\alpha_\mathrm{em}=e^2/(4\pi)$,
letting $T^a = {\rm diag}(\frac{2}{3},\frac{-1}{3},\frac{-1}{3},\dots)$ contain the quark electric charges,
the thermal photon production rate per unit volume of quark-gluon plasma can
be expressed as~\cite{McLerran:1985abc}
\beq
	\frac{\sd \Gamma_\gamma(k)}{\sd^3 k} = \frac{\alpha_\mathrm{em}}{\pi^2} \,
    \frac{\rho_\mathrm{V}(\om=k,k)}{4k} \, \frac{1}{\mathrm e^{k/T}-1} + \mathcal{O}(\alpha_\mathrm{em}^2),
	\label{eq:photonrate}
\eeq
in terms of the vector channel spectral function
$\rho_\mathrm{V}(\omega,k) = -\rho^{\mu}\,_{\mu}(\omega,{\bf k})$, where $k=|{\bf k}|$.
It~corresponds to the choice $\lambda = 1$ in the following linear combination, introduced in
Refs.~\cite{Brandt:2017vgl,Ce:2020tmx},
\beq
\rho(\om,k,\lambda) =
2 \rho_\mathrm{T}(\om,k) 
+ \lambda
\rho_\mathrm{L}(\om,k),
	\label{eq:rho_lambda}
\eeq
where
\beq
    \rho_\mathrm{T}(\om,k) = \frac{1}{2} \left(\delta_{ij} - \frac{k_i
    k_j}{k^2}\right) \rho_{ij} \quad {\rm{and}} \quad \rho_\mathrm{L}(\om,k) = \frac{k_i k_j}{k^2} \rho_{ij} - \rho_{00}
\eeq
denote the spatially transverse and longitudinal spectral functions, respectively.
As a consequence of current conservation, expression (\ref{eq:rho_lambda}) is independent
of $\lambda$ for light-cone kinematics $\om = k$.
Due to this fact, $\rho_\mathrm{V}(k,k)$ can be replaced by
$\rho(k,k,\lambda)$ with arbitrary $\lambda$ in the evaluation of the
photon rate~\cite{Brandt:2017vgl,Ce:2020tmx} of Eq.~(\ref{eq:photonrate}).
In Refs.~\cite{Brandt:2019shg,Ce:2020tmx}, the $\lambda = -2$ case 
was investigated, which corresponds to the difference of the transverse
and longitudinal channels
\beq \label{eq:TmL}
    \rho(\om,k,-2) = 2 \big[ \rho_\mathrm{T}(\om,k) - \rho_\mathrm{L}(\om,k) \big].
\eeq
This channel, $\rho(\om,k,-2)$, is particularly interesting 
because it is non-negative for $0\leq\om\leq k$, and highly suppressed when $\om > k$.
Therefore, it is very sensitive to infrared physics of interest.
It vanishes in the vacuum and also satisfies a superconvergent sum rule, 
demonstrated in Refs.~\cite{Brandt:2017vgl,Ce:2020tmx} and 
utilized as a constraint in the spectral reconstruction from 
Euclidean correlators in Ref.~\cite{Ce:2020tmx}.

In this work, we investigate the transverse channel, corresponding to 
using $\lambda=0$ in Eq.~(\ref{eq:rho_lambda}).
As we shall see below, it has complementary properties to the previously studied $\lambda=-2$ channel;
in particular, its spectral function is non-negative for all $\omega\geq 0$.
The transverse-channel Euclidean two-point functions of the vector current carrying a definite spatial momentum 
are related to their corresponding spectral function via (see e.g.~\!\cite{Meyer:2011gj})
\beq
    G_\mathrm{T}(\tau,k) = \int_0^\infty \, \frac{\sd\om}{2\pi} \, \rho_\mathrm{T}(\om, k) \, K(\om, \tau),
	\label{eq:sdecomp1}
\eeq
where the kernel is given as
\beq
	K(\om, \tau) = \frac{\cosh[\om(\beta/2-\tau)]}{\sinh(\om\beta/2)},
\eeq
and $\tau = {\rm i} t$.
This spectral representation of the Euclidean correlators will first be used
to confront them with theory predictions for the spectral function,
and, in a second stage, to fit an ansatz for the spectral function 
to the Euclidean correlators computed in lattice QCD.

We remark that in Sec.~\ref{sec:latres} and in Sec.~\ref{sec:estim} 
we use the flavour matrix $T^a = {\rm diag}(\frac{1}{\sqrt{2}},\frac{-1}{\sqrt{2}})$
in two-flavour QCD, i.e. we calculate the isovector vector current correlator.
This specifies in particular the normalisation of our results for $G_{\rm T}$, $G_{00}$ or $\rho_{\rm T}$.
If we are willing to use the approximate SU(3) flavour symmetry in the
high-temperature phase of QCD and neglect dynamical strange-quark effects as well as the charm contribution,
the computed correlators simply need to be multiplied by the charge factor,
$C_\mathrm{em} = (2/3)^2+(-1/3)^2+(-1/3)^2 = 2/3$, to obtain the electromagnetic current
correlator in the physical QGP.

\subsection{Hydrodynamics}
\label{sec:hydro}

The long-wavelength behavior of the spectral function can be studied with
the help of hydrodynamics.
Let $D$ be the diffusion coefficient for the conserved electric charges.
In the hydrodynamic regime $\om, k \ll D^{-1}$, the first-order
hydrodynamic prediction for the transverse channel spectral function is%
~\cite{Hong:2010at}
\beq
    \frac{\rho_\mathrm{T}(\om,k)}{\om} \approx 2 D \chi_\mathrm{s},
	\label{eq:rhoT,hydro}
\eeq
where $\chi_\mathrm{s}$ denotes the static quark susceptibility,
\beq
	\chi_\mathrm{s} = \int_0^\beta \sd \tau \int \sd^3 x \,\langle J_0(\tau, {\bf x}) J_0(0, {\bf 0}) \rangle.
\eeq
The diffusion coefficient can be expressed using the electrical conductivity, $\sigma$,
as $D=\sigma/\chi_\mathrm{s}$.

The functional form of $\rho_\mathrm{T}$ in Eq.~(\ref{eq:rhoT,hydro}) reveals
the advantage of investigating this channel, namely that it does not
couple to the diffusion pole in contrast to $\rho_\mathrm{V}$ or
$\rho(\om,k,-2)$~\cite{Ghiglieri:2016tvj,Ce:2020tmx}.  Therefore
$\rho_\mathrm{T}(\om,k)/\om$ around $\omega=0$ does not contain any peak-like
structure of arbitrarily small width as $k\to 0$.\footnote{The only exception to this statement is for a plasma
of strictly non-interacting particles, for which hydrodynamic
predictions do not apply.}  At extremely high temperatures in QCD, which
implies a small value of the strong coupling constant $g$, a kinetic theory
treatment eventually becomes applicable and a narrow peak of width $\sim g^4 T$
appears~\cite{Hong:2010at}.

Following Ref.~\cite{Ghiglieri:2016tvj}, we define the effective
diffusion coefficient
\beq
D_{\mathrm{eff}}(k) \equiv \frac{\rho_\mathrm{T}(\om=k,k)}{2 \chi_s k},
	\label{eq:TDeff}
        \eeq
which tends to $D$ as $k \to 0$ and is the key dynamical ingredient to evaluate the
thermal photon production rate of Eq. (\ref{eq:photonrate}).

\subsection{Resummed spectral functions at NLO in thermal QCD}
\label{sec:pert}

Analytic predictions are also available for the spectral functions
in the weak-coupling limit.
The spectral function in the timelike regime --- which is relevant for
dilepton production --- has been investigated using perturbative calculations
since the seminal works of Refs.~\cite{McLerran:1985abc,Baier:1988xv}.
In most of these perturbative calculations, the studied channel
was the vector channel spectral function, $\rho_\mathrm{V}$.
Focusing on the transverse channel, leading-order (LO) results (corresponding to non-interacting quarks)
are available for massless quarks in e.g.\ Refs.~\cite{Aarts:2005hg,Laine:2013vma,Ghisoiu:2014mha}.
Recently, this has been extended to NLO at $\mathrm{O}(g^2)$
calculating the perturbative contributions up to two-loops.
The details of this impressive two-loop calculation --- valid both for spacelike and timelike
virtualities --- can be found in Ref.~\cite{Jackson:2019mop}, while a comparison to lattice correlators
has been presented in Ref.~\cite{Jackson:2019yao}.

As we discussed in Sec.~\ref{sec:basics}, the relevant information for the 
thermal photon production is coming from the spectral function determined at the
light-cone.
The thermal photon emission rate vanishes for non-interacting quarks~\cite{Laine:2013vma}.
The NLO perturbative thermal photon emission rate can be either
determined by evaluating the NLO spectral functions at the light-cone
or by making use of the computation in Ref.~\cite{Arnold:2001ms}.
The $\mathrm{O}(g^2)$ calculation of Ref.~\cite{Arnold:2001ms} has been
extended to $\mathrm{O}(g^3)$ by taking into account contributions from soft gluons~\cite{Ghiglieri:2013gia}.

The strict two-loop perturbative spectral function
develops a logarithmic singularity at $\om=k$~\cite{Jackson:2019mop,Jackson:2019yao}.
It originates from multiple rescatterings of a quark taking part off-shell
in the inelastic annihilation process that produces a photon
or in bremsstrahlung~\cite{Aurenche:2000gf,Arnold:2001ms,Aurenche:2002wq,Carrington:2007gt,Ghisoiu:2014mha}.
This infrared (IR) singularity is also present in the NLO calculation of the
real photon rate~\cite{Aurenche:2000gf,Arnold:2001ms}.
The effect is called the Landau-Pomeranchuk-Migdal (LPM) effect~\cite{Aurenche:2000gf,Arnold:2001ms},
and can be handled by implementing a proper resummation of ladder diagrams,
called the LPM resummation~\cite{Arnold:2001ms,Aurenche:2002wq,Ghisoiu:2014mha,Jackson:2019yao}.

In order to compare lattice and perturbation theory results,
we used the publicly-available implementation of the two-loop calculation
presented in Ref.~\cite{Jackson:2019mop} and computed
the LPM resummation based on Refs.~\cite{Ghisoiu:2014mha,Jackson:2019yao}.
In our implementation, we used a window function to restrict the LPM
contribution to frequencies around the light-cone.
The NLO perturbative spectral function complemented by the LPM
contribution is called NLO+LPM in the following.
For the spatial momentum of $k=\pi T$, we display this spectral function in
Figure~\ref{fig:pipoly,rho} using dashed lines.

\subsection{\texorpdfstring{$\mathcal{N}=4$}{𝒩=4} supersymmetric Yang-Mills theory}
\label{sec:adscft}

In the $\mathcal{N}=4$ supersymmetric Yang-Mills (SYM) theory,
the thermal spectral functions can  be obtained analytically not only in the
weak coupling limit, but in the strong coupling (and large-$N_\mathrm{c}$) regime as well~\cite{Caron-Huot:2006pee}.
The field content of the theory is an SU($N_\mathrm{c}$) gauge field together with massless scalar and fermionic fields
in the adjoint representation of the gauge group.
In the limit of infinite 't Hooft coupling and infinite number of colours,
one can determine the spectral function by making use of the AdS/CFT
correspondence and then numerically solving an ordinary differential
equation.
Our primary interest, the spectral function in the transverse channel,
is a smooth function having a similar asymptotic behavior --- proportional
to $\om^2$ --- at high frequencies as the transverse spectral function
in thermal QCD.
Although the spectral functions in SYM in the strong
coupling limit cannot quantitatively describe
those of thermal QCD, their qualitative features are nevertheless instructive
and may well be of relevance at $T\simeq 250$\,MeV.
An example of such a spectral function at strong coupling is shown in Fig.~\ref{fig:pipoly,rho}
for a spatial momentum $k=\pi T$ with a solid line.

\section{Lattice computation of the transverse correlators}
\label{sec:latres}
\subsection{Ensembles and statistics}
\label{sec:stat}

We employ dynamical O($a$)-improved Wilson fermions to simulate two degenerate flavours of quarks
with an \emph{in vacuo} pion mass around 270 MeV.
The details of the lattice action can be found in Ref.~\cite{Fritzsch:2012wq} and references therein.
The temperature is $T \simeq 250$ MeV, well above the transition 
temperature, estimated to be about $T_\mathrm{c}\simeq211$ MeV in $N_\mathrm{f}=2$ QCD~\cite{Brandt:2016daq}.
We use the ensembles already presented in Ref.~\cite{Ce:2020tmx}; cf.\ Table~I in the latter reference.
Although the pion mass is larger than its physical value on these ensembles,
we emphasize that quark-mass effects on the correlator are suppressed
by $(m/T)^2$ in the chirally symmetric phase.
For our two coarsest ensembles, labeled as F7 and O7, the bare parameters
have been set identical to the vacuum F7 and O7 ensembles used in
Ref.~\cite{Fritzsch:2012wq,Engel:2014cka}, while for the finer ensembles, labeled as W7 and X7, the tuning
to the line of constant physics was performed in Ref.~\cite{Steinberg:2021bgr}.
The lattice spacings for these ensembles are around $a\simeq 0.066$, 0.049, 0.039
and 0.033 fm with an error of about 1\%~\cite{Engel:2014cka}.
The physical volume for all ensembles is around $L\simeq 3.1$ fm.
\subsection{Lattice observables}
\label{sec:obs}

In this section, we introduce the various discretized lattice correlators
that we used to perform the continuum extrapolation in Sec. \ref{sec:contlim}.
We consider the two-point function of the isovector vector current in QCD with exact
isospin symmetry. This allows for precise comparisons with weak-coupling predictions
in $N_{\rm f}=2$ QCD.
As discussed at the end of Sec.~\ref{sec:basics},
the correlator of the electromagnetic current in the physical QGP
can be obtained approximately by multiplying our correlator
by the charge factor $C_\mathrm{em} =  2/3$.

The bare local and the conserved vector current are defined as
\beq
	V_\mu^\mathrm{L}(x) = \widebar \Psi(x)\frac{\tau_3}{\sqrt{2}}\gamma_\mu\Psi(x),
\eeq
and
\beq
	V_\mu^\mathrm{C}(x) = \frac{1}{2} \left[ \widebar{\Psi}(x+a\hat\mu) (1 + \gamma_\mu) U^\dag_\mu(x) \frac{\tau_3}{\sqrt{2}}\Psi(x) - \widebar{\Psi}(x)(1 - \gamma_\mu) U_\mu(x)\frac{\tau_3}{\sqrt{2}} \Psi(x+a\hat\mu) \right],
\eeq
respectively, where $\Psi=(u,d)^\top$ represents the isospin doublet of 
mass-degenerate quark fields and $\tau_3$ is the diagonal Pauli matrix.
As in Ref.~\!\cite{Ce:2020tmx}, we have not implemented the additive O($a$) improvement
of vector currents, as the contribution of the improvement terms to the two-point functions
would be suppressed by the quark mass in the chirally restored phase.
Using the currents above, we define the following bare unimproved correlators
\begin{align}
	G_{\mu\nu}^\mathrm{LL}(\tau, {\bf k}) &= a^3 \sum_{\bf x} \, \mathrm e^{\mathrm i {\bf k \cdot x}} \langle V_{\mu}^\mathrm{L}(\tau, {\bf x}) V_{\nu}^\mathrm{L}(0, {\bf 0})^\dagger \rangle,
\label{eq:GmunuLL}\\
	G_{\mu\nu}^\mathrm{CC}(\tau+a\delta_{\mu 0}/2-a\delta_{\nu 0}/2, {\bf k}) &= a^3 \sum_{\bf x} \, \mathrm e^{\mathrm i {\bf k \cdot} ({\bf x} + a\hat{\mu}/2 - a\hat{\nu}/2)} \langle V_{\mu}^\mathrm{C}(\tau, {\bf x}) V_{\nu}^\mathrm{C}(0, {\bf 0})^\dagger \rangle,
\label{eq:GmunuCC}\\
	G_{\mu\nu}^\mathrm{LC}(\tau-a\delta_{\nu 0}/2, {\bf k}) &= a^3 \sum_{\bf x} \, \mathrm e^{\mathrm i {\bf k \cdot} ({\bf x} - a\hat{\nu}/2)} \langle V_{\mu}^\mathrm{L}(\tau, {\bf x}) V_{\nu}^\mathrm{C}(0, {\bf 0})^\dagger \rangle,
\label{eq:GmunuLC}\\
	G_{\mu\nu}^\mathrm{CL}(\tau+a\delta_{\mu 0}/2, {\bf k}) &= a^3 \sum_{\bf x} \, \mathrm e^{\mathrm i {\bf k \cdot} ({\bf x} + a\hat{\mu}/2)} \langle V_{\mu}^\mathrm{C}(\tau, {\bf x}) V_{\nu}^\mathrm{L}(0, {\bf 0})^\dagger \rangle.
\label{eq:GmunuCL}
\end{align}
Under time reflections the local-conserved correlator is transformed
into the conserved-local one, $G_{\mu\nu}^\mathrm{CL}(\tau, {\bf k}) \overset{T}{\to} G_{\nu\mu}^\mathrm{LC}(-\tau, {\bf k})$,
and vice versa.
Using this fact, we averaged the two appropriately, and we refer 
to it with the the superscript $\mathrm{LC}$ in the following.
For the transverse correlator we need only the spatial components of the 
correlators, which are defined on the lattice sites according to Eqs.~(\ref{eq:GmunuLL}-\ref{eq:GmunuCL}).
In the case of local-conserved or conserved-local correlators,
however, we note that the charge-charge correlator can be evaluated on site $\tau$ 
by averaging $G_{00}^\mathrm{LC}(\tau+a/2,{\bf k})$ and $G_{00}^\mathrm{LC}(\tau-a/2,{\bf k})$.

To obtain the correlator in the transverse channel, we first evaluated
\beq
\hat G_\mathrm{T}^{\alpha}(\tau,{\bf k}) = \frac{1}{2} \sum_{i,j=1}^3 \Big( \delta_{ij} - \frac{k_i k_j}{k^2}  \Big) G_{ij}^{\,\alpha}(\tau,{\bf k}),
\qquad\alpha=\mathrm{LL,LC,CC}.
	\label{eq:GTdef}
\eeq
Imposing time-reversal symmetry and translation invariance, we then
symmetrized the correlators in the time direction and averaged over the momentum orientations:
\beq
G_\mathrm{T}^{\,\alpha}(\tau, k) = \frac{1}{2N_k} \sum_{|{\bf k^\prime}|=k}  \left( \hat G_\mathrm{T}^{\alpha}(\tau, {\bf k}') + \hat G_\mathrm{T}^{\alpha}(\beta - \tau, {\bf k}') \right),
\eeq
where $N_k$ is the number of momenta of norm $k$.

\subsection{Continuum extrapolation}
\label{sec:contlim}

For the continuum extrapolation, we first interpolated the lattice correlators
to the time separations which correspond to our finest lattice, X7.
We applied two interpolation methods, Akima and monotonic cubic spline~\cite{Akima:1970abc,Steffen:1990abc,GSL:2019}.
For the interpolation, we normalized $G_\mathrm{T}/G_{00}$ by the LO continuum
transverse correlator, i.e.\ we multiplied by $T^3/G_\mathrm{T}^\mathrm{LO}$
in order to have a flattened interpolant.
After the interpolation we removed this factor.%
\footnote{Alternatively,
one can carry out the tree-level improvement of the data 
at this step, i.e.\ before the interpolation.}
In order to avoid the renormalization of the local currents, we divided
the bare correlator by the bare static susceptibility 
computed using the same discretization
\beq
\widebar{G}^{\alpha}_\mathrm{T} \equiv \frac{G^{\alpha}_\mathrm{T}(\tau,k)}{\chi_\mathrm{s}^{\alpha}(\tau) T},
	\qquad
    \chi_\mathrm{s}^{\alpha}(\tau) = \beta G^{\alpha}_{00}(\tau,0).
\eeq
In the following, we omit the label $\mathrm T$ denoting the transverse channel.

For the continuum extrapolation, we carried out fits by using only a single 
discretization or multiple discretizations simultaneously of the transverse current 
correlators for each $\tau$ and $k$.
When using multiple discretizations, we performed constrained correlated fits
using the following fit ansatz
\beq
d(a/\beta,{\bf c}^{\alpha}) = c_0 + c_1^{\alpha} \,(a/\beta)^2 + c_2^{\alpha}\, (a/\beta)^4,
\eeq
where $c_0$ is the estimate of the continuum limit from a particular fit,
and the $c_1^{\alpha}$ and $c_2^{\alpha}$ parameters are characterizing the approach
to the continuum of the different discretized correlators.
We took into account the correlations between the different discretizations,
and minimized the chi-squared statistic
\beq
\chi^2 = \sum_{e=1}^{N_\mathrm{e}} \sum_{\alpha,\alpha^{\prime}} \Big[ \widebar{G}^{\alpha}(a_e/\beta) - d(a_e/\beta, {\bf c}^\alpha) \Big]
\mathrm{Cov}^{-1}_{e,\alpha\alpha^{\prime}} \Big[ \widebar{G}^{\alpha^\prime}(a_e/\beta) - d(a_e/\beta, {\bf c}^{\alpha^\prime})\Big],
\eeq
where the index $e$ runs over the ensembles and $\alpha,\alpha^\prime = \mathrm{LL, LC, CC}$.
A regularized covariance matrix $\mathrm{Cov}$ was obtained by multiplying the off-diagonal
elements by 0.95.
The regularization had a non-negligible effect only in the case of linear fits
(i.e.\ when setting the $c_2^{\alpha}$ coefficients to zero),
for which it led to an increased number of acceptable fits.

To increase the robustness of the continuum limit, we implemented a multiplicative 
tree-level improvement of the lattice data
\beq
	\widebar{G}^{\alpha}_\mathrm{TLI}(\tau,k) \equiv \widebar{G}^{\alpha}(\tau,k)
	\left[\frac{\widebar{G}(\tau,k)}{\widebar{G}^{\alpha}(\tau,k)}\right]_\mathrm{LO},
\eeq
where $\widebar{G}^{\alpha}_\mathrm{TLI}$ is the tree-level improved correlator,
$\widebar{G}_\mathrm{LO}$ and $\widebar{G}^{\alpha}_\mathrm{LO}$ are the continuum 
and the lattice leading-order perturbation theory results, respectively.
The tree-level improvement reduces the difference between the various
discretizations at finite lattice spacing, which results in more fits having
acceptable $p$-values.
It reduces the continuum extrapolated value at smaller Euclidean time
separations, $\tau T<0.25$.
The effect of the improvement turned out to be much milder
--- almost negligible on the final continuum result --- for larger distances, $\tau T \ge 0.25$.

\begin{figure}[t!]
	\includegraphics[scale=0.64]{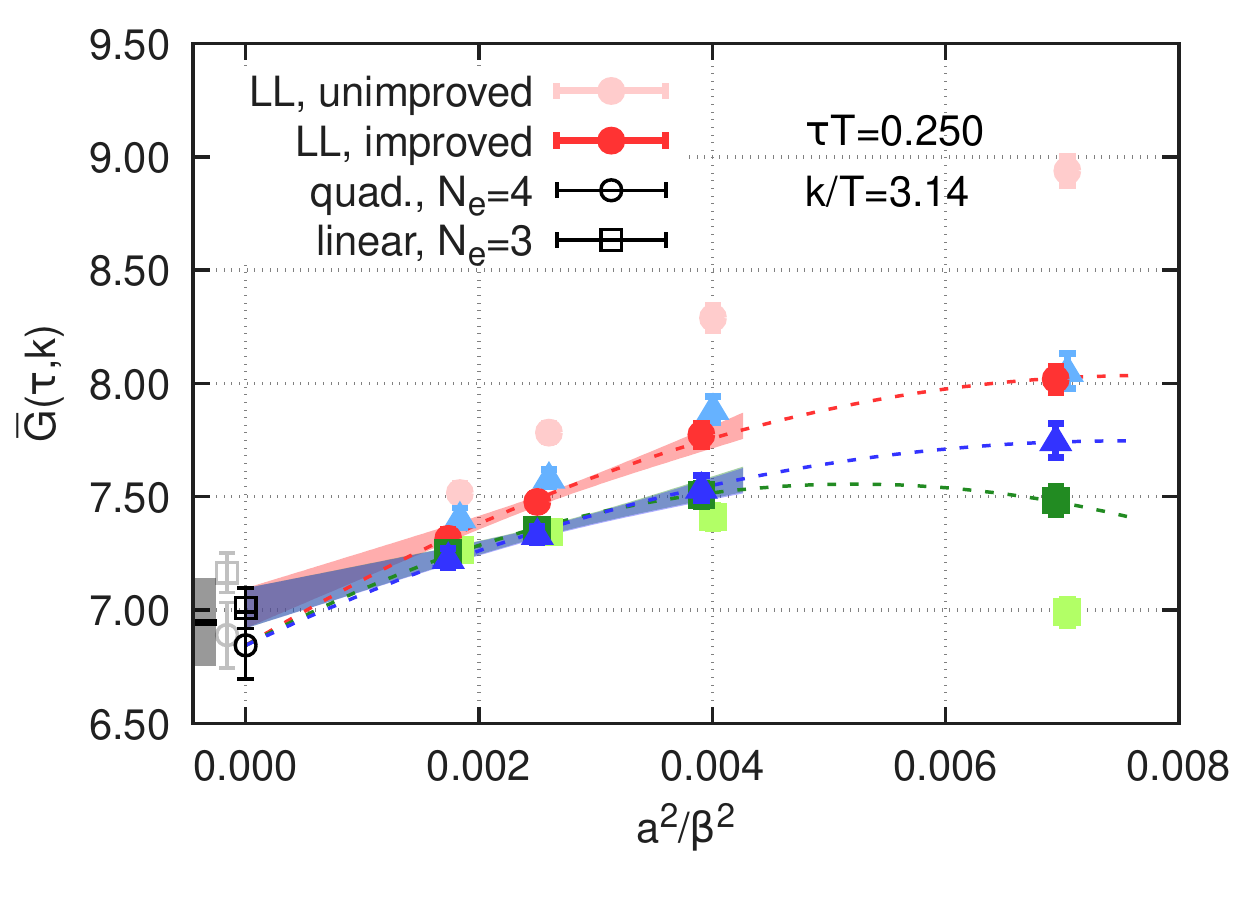}
	\includegraphics[scale=0.64]{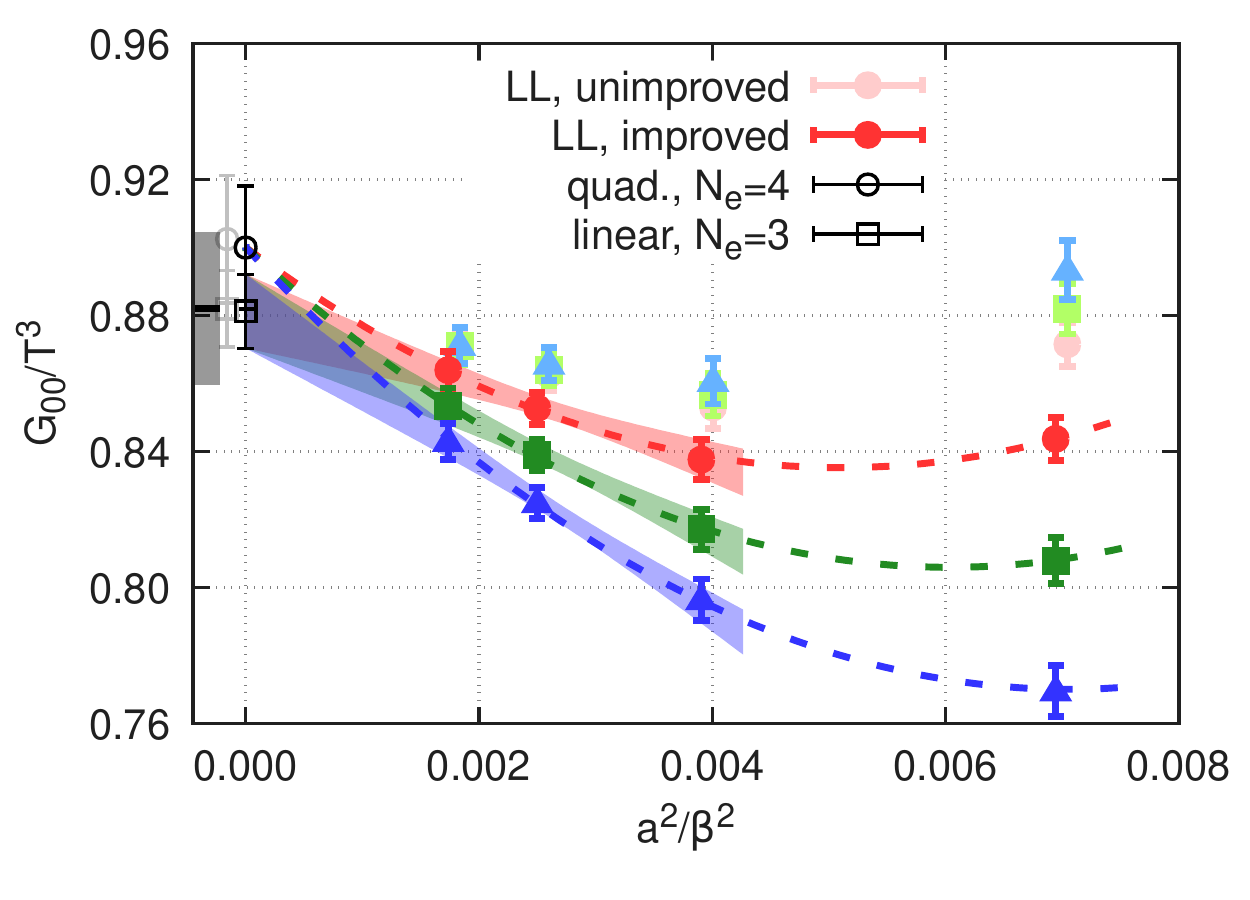}
\caption{
{\bf Left panel}: Simultaneous continuum extrapolation at $\tau T= 0.25$ of the local-local (LL),
local-conserved (LC) and conserved-conserved (CC) tree-level improved correlators plotted 
with red circles, blue triangles and green squares, respectively.
The unimproved correlator values are shown with fainter symbols.
The coloured bands show the linear fit using the three finest lattice spacings and all
discretizations, while the dashed lines illustrates the quadratic fit using all data points.
The results of the linear (quadratic) extrapolations are shown as empty squares (circles).
The fainter, shifted, grey symbols show the continuum values obtained using the unimproved correlators.
The leftmost grey band represents our final continuum estimate, including its systematic error.
{\bf Right panel}: Simultaneous continuum extrapolation of the isovector charge-charge correlator
required to compute the static susceptibility.
The colour code is the same as on the left panel.
}
\label{fig:contlim}
\end{figure}

In order to estimate the systematic errors, we carried out extrapolations 
using the tree-level improved as well as the unimproved data.
Further systematic variations included using two interpolation methods
to interpolate to the same $\tau T$ points on all ensembles and varying 
the number of data points used in the fits.
For the linear fits, only the three finest ensembles have been used and we 
investigated the robustness of the results by omitting one or two data points
belonging to the coarsest two out of three ensembles.
In the case of quadratic fits, we proceeded similarly, but we omitted one
or two data points from the coarsest three out of four ensembles.
When we left out more than one point, we always discarded them from the same
discretization.
Furthermore, we carried out extrapolations by using only one or two 
discretizations.
These various changes in the analysis resulted in a total number of 104 fits.
The $\chi^2$ values, the number of fit parameters and the number of data
points have been used to calculate the Akaike Information Criterion (AIC)
weight of each fit~\cite{Akaike:1973abc,Borsanyi:2020mff}.
Using these weights, we built a histogram and quoted the median as our final
continuum result.
The histogram is used in the later stages of the analysis.

We also perform the continuum limit of the static susceptibility, which requires
including the $Z_{\mathrm V}$ renormalization constant when the local
discretization of the current is used.
To obtain the values of $Z_{\mathrm V}$ at the bare couplings of our ensembles,
we utilized the parametrization of $Z_{\mathrm V}(g_0^2)$ given in Ref.~\cite{DellaMorte:2005xgj}.
Then, adopting a similar procedure to determine the continuum limit of
the static susceptibility as we used for the transverse correlator, we obtain 
$\chi_\mathrm{s}/T^2= \beta^3 G_{00}(\beta/2,0) = 0.882(11)_{\mathrm{stat}}(19)_{\mathrm{sys}}$,
see the right panel of Fig.~\ref{fig:contlim}.
Compared with our result of Ref.~\cite{Ce:2020tmx}, we quote a more
conservative systematic error due to the continuum limit primarily due to the
inclusion of quadratic fits in $a^2$ in this work.
We have also investigated that the use of other determinations of the
renormalization factor~\cite{DallaBrida:2018tpn} and the inclusion of the
mass-dependent improvement, which lead to sub-leading differences compared to
the systematic error from the continuum extrapolation.

In order to obtain $G_\mathrm{T}/T^3$ in the continuum, we multiplied the continuum limit of $G_\mathrm{T}/G_{00}$
by the continuum estimate of $G_{00}/T^3$.
The transverse correlator divided by the charge-charge correlator
is shown in the left panel of Fig.~\ref{fig:GT,comparison} for $k/T\approx 4.97$, and in units of temperature in the right panel.
The statistical error on $G_\mathrm{T}/T^3$ is typically around $0.25 - 0.6\%$, while
it is in the range $1.1 - 1.4\%$ for $G_\mathrm{T}/G_{00}$.
\subsection{Continuum-extrapolated correlators vs.\ theory predictions}
Having obtained the transverse-channel correlators in the continuum, we compare
them to various theory predictions.
In this comparison, we neglect the $O((m/T)^2)$ quark-mass effects
present in the lattice results, since $m/T\approx 0.05$ in our simulations.

We find that the LO correlator is about $5-20\%$ larger
at Euclidean time separations $\tau T > 0.4$ for momenta $k/T \gtrsim 3.85$;
see Fig.~\ref{fig:GT,comparison}.
The deviation is smaller, around $\sim$ 5\%, for smaller momenta.
It reduces towards smaller time separations, and below a certain ($k$-dependent) $\tau T$ value,
the LO correlator is smaller by about 5\% than our continuum result.
The NLO+LPM correlator, however, is only about
$3-4\%$ larger than the lattice result for all momenta and for all 
time separations  $0.17 \lesssim \tau T \le 0.5$ for which we could reliably determine the continuum limit.
We conclude that the perturbative corrections to the LO correlator noticeably improve its agreement
with our lattice correlator.
The strongly coupled $\mathcal{N}=4$ SYM theory result also provides
a relatively good description, even though that result concerns a different non-Abelian gauge theory
and applies in the large-$N_c$ limit.
The lower panels of Fig.~\ref{fig:GT,comparison} show the ratio of the
perturbative and the SYM results to the lattice results.

\begin{figure}[t!]
	\includegraphics[scale=0.64]{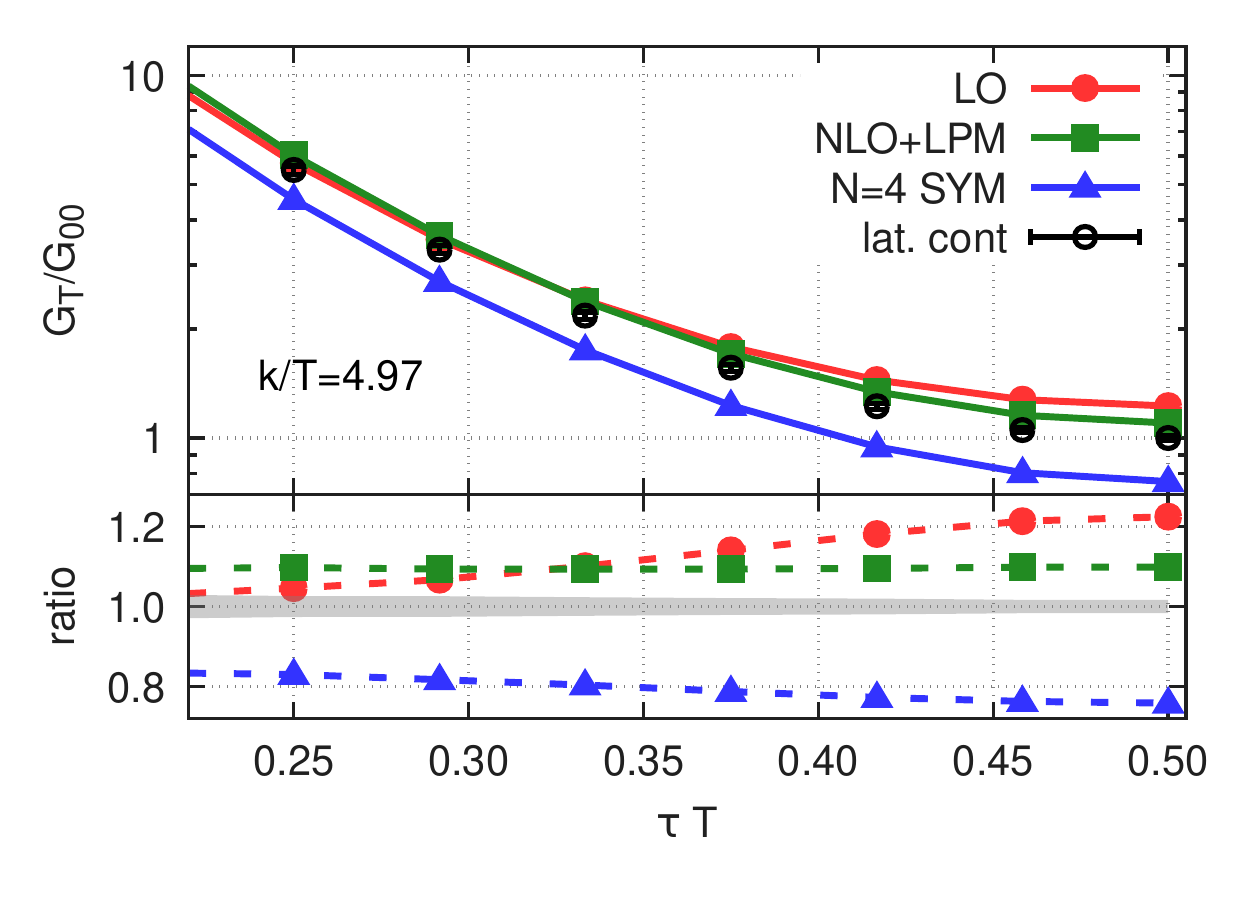}
	\includegraphics[scale=0.64]{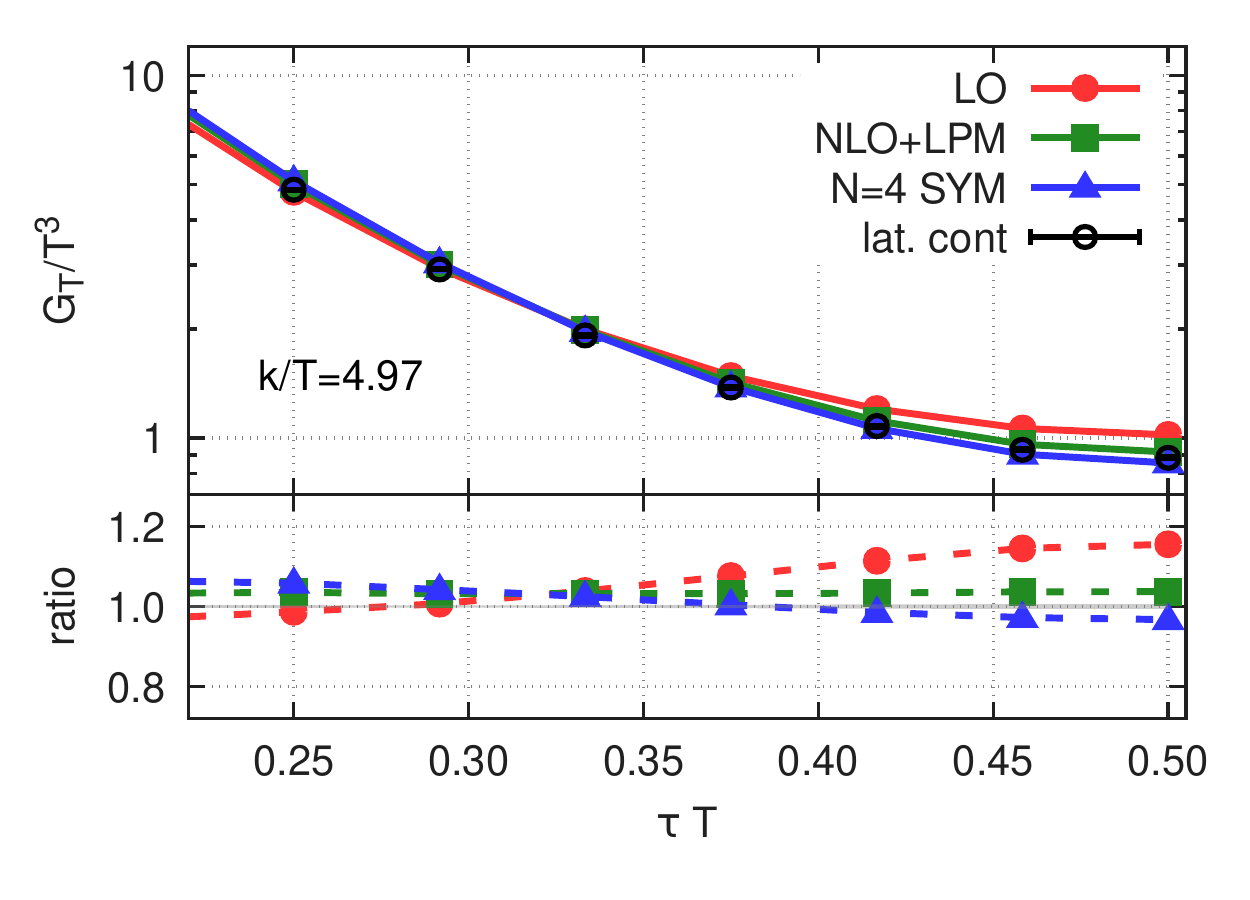}
\caption{
    {\bf Left panel}: Comparison of the $G_\mathrm{T}/G_{00}$ correlators.
The bottom panels show the correlators normalized by the continuum correlator.
The colour code is as in the top panel.
The NLO+LPM result was obtained by using $\alpha_\mathrm{s} = 0.25$.
{\bf Right panel}: Comparison of $G_\mathrm{T}/T^3$.
We set $N_\mathrm{c}=3$ to obtain the SYM theory result.
\label{fig:GT,comparison}
}
\end{figure}

A word on the multiplicative normalization of the theory predictions is in order.
For the perturbative predictions, while no prescription is needed for $G(\tau,k)$,
a choice must be make to obtain $\bar G(\tau,k)$. Here we have normalized
both the LO and the NLO+LPM correlators by the $\mathrm{O}(g^6 \ln g)$
susceptibility~\cite{Vuorinen:2002ue}.
In Ref.~\cite{Vuorinen:2002ue}, the coefficient of one of the undetermined
$\mathrm{O}(g^6)$ term was estimated to be $C(N_\mathrm{f}=2)\approx -45$,
which results in $\chi_{\mathrm{pert}}/T^2 \approx 0.83$, the value which we
used for Fig.~\ref{fig:GT,comparison}.
This value, however, is 6\% lower than the continuum extrapolated value we obtained.
Based on our lattice continuum result, $C(N_\mathrm{f}=2)\approx 33$ would result
in a better agreement for $\chi_\mathrm{s}/T^2$.
However, we note that the contributions from successive higher orders in the perturbative
calculation of $\chi_{\mathrm{pert}}/T^2$ are similar in size, so the estimation
for $C(N_\mathrm{f}=2)$ is to be treated with caution.
As for the strongly coupled, large-$N_c$ $\mathcal{N}=4$ SYM correlator, we note that $\bar G(\tau,k)$ is a natural 
quantity to compare across different thermal systems; in particular, the dependence on the
number of colours drops out. 
In order to compare the correlator $G(\tau,k)$ itself, a certain choice must be made for the susceptibility.
We have chosen to set  $\chi_\mathrm{SYM}/T^2=N_\mathrm{c}^2/8=9/8$.

\section{Analysis of the transverse-channel spectral function}
\label{sec:estim}

Given the transverse-channel Euclidean correlation function in the continuum limit, 
we proceed to analyze the corresponding spectral function via Eq.~\!(\ref{eq:sdecomp1}).
Here, we present an analysis based directly on $G_\mathrm{T}/T^3$ in Secs.~\ref{sec:BG} and \ref{sec:fitmethod},
since we found the correlator to be statistically particularly precise.
In addition, we have seen that the NLO+LPM correlator is only a few percent off the lattice correlator,
which encourages us to use that prediction as a baseline in Sec.~\!\ref{sec:fitmethod}.
In contrast, since the primary continuum-extrapolated quantity was $\widebar{G} = G_\mathrm{T}/G_{00}$,
Ref.~\!\cite{Torok:2021ujr} contains an analysis based on that ratio.

We start in Sec.~\!\ref{sec:BG} by using the Backus-Gilbert method in order to perform the comparison
between lattice data and theory predictions in frequency space without introducing
any model dependence.
Sec.~\!\ref{sec:fitmethod} then presents fits to the lattice data in order to  determine the spectral function,
with a particular focus on lightlike kinematics, $\omega=k$.
For simplicity of notation, we omit the spatial momentum from the
arguments in the following sections.

\subsection{Backus-Gilbert method: smeared spectral functions vs.\ theory predictions}
\label{sec:BG}

The Backus-Gilbert method is a model-independent approach to overcome
the spectral reconstruction problem~\cite{Backus:1968abc}.
By using this method one can determine a local average of the
spectral function around a given value of the frequency.
In the present case it is also favourable to introduce a rescaling 
function, $f(\om)$, which in particular removes the singularity at vanishing
frequency of the kernel $K(\om, \tau)$.
Thus the new kernel is defined as
\beq
	K_f(\om, \tau) \equiv K(\om, \tau) f(\om),
\eeq
with the rescaling function $f(\om)$ being specified later,
but satisfying $f(\om) \propto \om$ as $\om \to 0$.
The smeared, rescaled spectral function, $\hat{\rho}(\om)/f(\om)$,
is defined in this case as
\beq
	\frac{\hat{\rho}(\om)}{f(\om)} \equiv
	\int_0^\infty \sd \om^\prime \Delta(\om, \om^\prime) \frac{\rho(\om^\prime)}{f(\om^\prime)},
	\label{eq:rhohatf}
\eeq
where $\Delta(\om,\om^\prime)$ is the so-called resolution function or
averaging kernel, which is completely specified in terms of some 
coefficients, $g_i(\om)$, and is given as
\beq
	\Delta(\om, \om^\prime) = \sum_i g_i(\om) K_f(\om^\prime, \tau_i).
	\label{eq:Delta}
\eeq
Inserting back $\Delta(\om,\om^\prime)$ of Eq.~(\ref{eq:Delta})
into Eq.~(\ref{eq:rhohatf}), we find that the filtered spectral
function is the linear combination of the Euclidean correlator data,
\beq
	\hat{\rho}(\om) = f(\om) \sum_i g_i(\om) G_{\rm{T}}(\tau_i).
	\label{eq:rhohat}
\eeq

\begin{figure}[t!]
	\includegraphics[scale=0.64]{{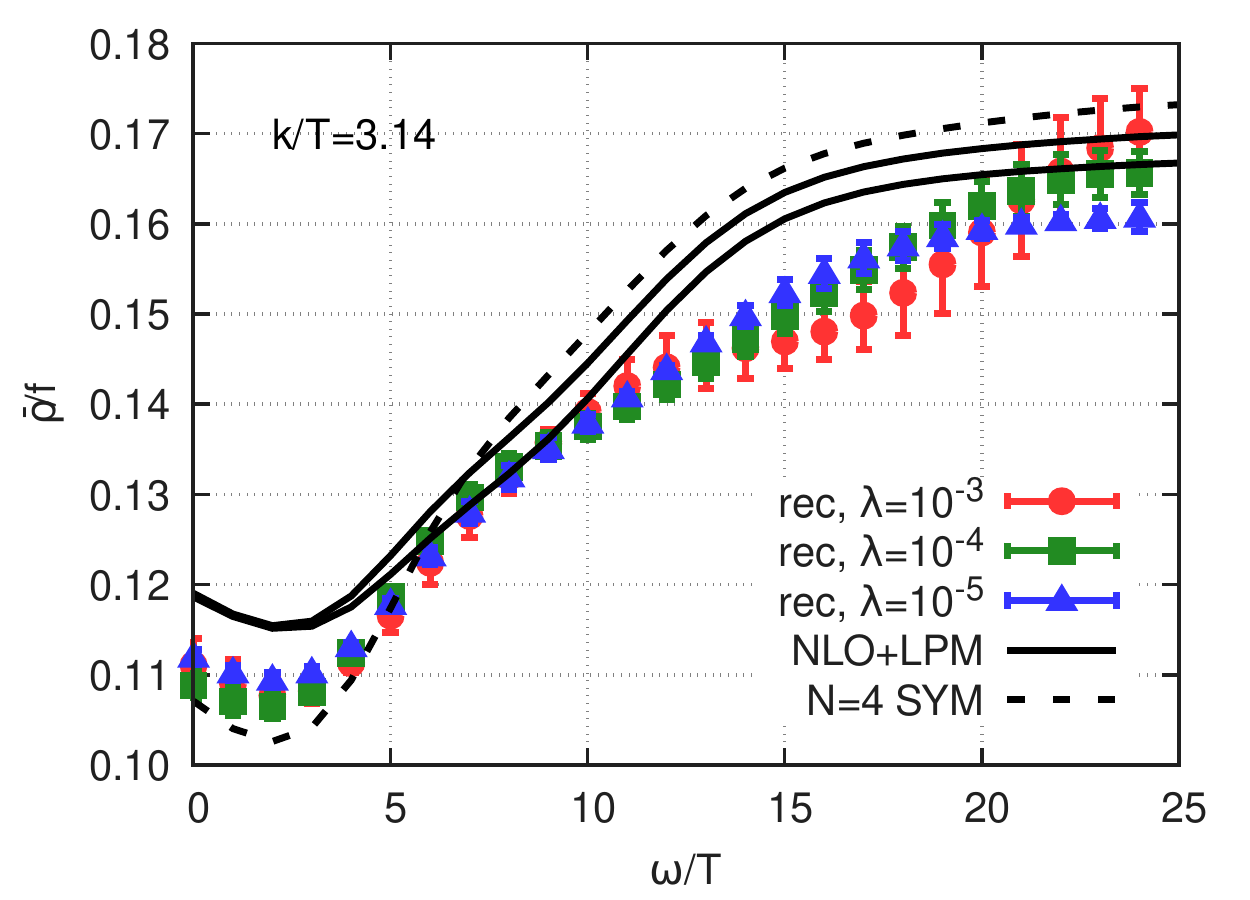}}
	\includegraphics[scale=0.64]{{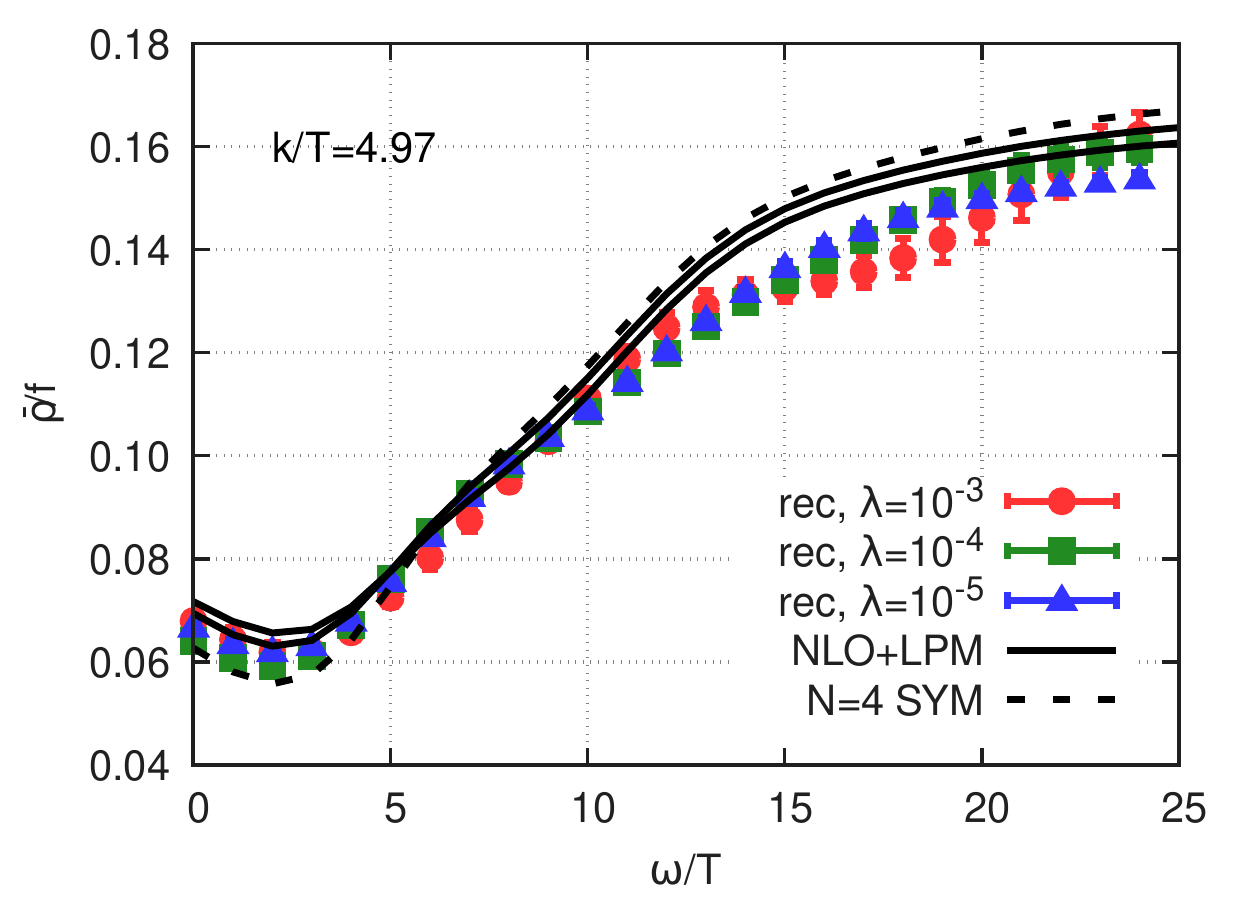}}
\caption
{
	\label{fig:BG,rhohat}
	Comparison of filtered, rescaled, transverse channel spectral functions obtained
	using the continuum lattice data (with colours) and the perturbative spectral 
	function (continuous, black) as well as the $\mathcal{N}=4$ SYM
	theory (dashed, black) for $k/T \approx 3.14$ and $k/T \approx 4.97$,
	left and right panel, respectively.
	The renormalization scale was set to $\mu= 2\pi T$ or $\mu= 3\pi T$
	when using the NLO+LPM spectral function, which results in two curves.
    The renormalization scales $2\pi T$ and $3\pi T$ correspond to $\alpha_\mathrm{s}\simeq 0.31$
and $\alpha_\mathrm{s}\simeq 0.25$, respectively.
	Errors are only statistical.
}
\end{figure}

\noindent
The coefficients, $g_i(\om)$, are determined in the Backus-Gilbert
method by minimizing the second moment of the squared resolution function
\beq
	A[g] \equiv \int_0^\infty \sd \om^\prime \, (\om - \om^\prime)^2 [\Delta(\om,\om^\prime)]^2 = \sum_{i,j} g_i(\om) A_{ij}(\om) g_j(\om)
\eeq
subject to the constraint
\beq
	\int_0^\infty \sd \om^\prime \Delta(\om, \om^\prime) = 1.
\eeq
This minimization ensures that the width of resolution function is
as small as possible and that, at fixed $\omega$, $\Delta(\omega,\omega')$ has unit area.
The minimizing solution is given as
\beq
	g_i(\om) = \frac{A_{ij}^{-1}(\om)R_j}{R_k A_{kl}^{-1}(\om)R_l},
	\label{eq:gi}
\eeq
where
\beq
	A_{ij}(\om) = \int_0^\infty \sd \om^\prime K_f(\om^\prime, \tau_i) K_f(\om^\prime, \tau_j) (\om - \om^\prime)^2 \quad {\mathrm{and}} \quad R_i = \int_0^\infty \sd \om K_f(\om, \tau_i).
\eeq
The matrix ${\bf A}$ is ill-conditioned in practice,
therefore an error functional is added to $A[g]$ which
serves as a regulator.
As a consequence, ${\bf A}$ has to be replaced by ${\bf A}^{\mathrm{reg}}$
in Eq. (\ref{eq:gi}), where
\beq
A^{\mathrm{reg}}_{ij}(\om) = \lambda A_{ij}(\om) + (1-\lambda) \Cov_{ij}[G_\mathrm{T}],
	\label{eq:Aijreg}
\eeq
and $0 \le \lambda \le 1$ is the regularization parameter 
compromising between stability and re\-so\-lution.
The smaller the value of $\lambda$, the larger the regularization.
In Eq. (\ref{eq:Aijreg}), $\Cov[G_\mathrm{T}]$ stands for the covariance matrix
of the Euclidean correlator.
We note that in our numerical implementation we worked in the units
of temperature.

We emphasize again that from the filtered spectral function one cannot model-independently
determine the value of $\rho(\om)$ itself, but it could be useful to compare
it to a similarly smoothened spectral function coming from other approaches.
Once the $g_i(\om)$ coefficients are determined via Eq.~(\ref{eq:gi}) by
utilizing the covariance matrix of the data, the same resolution function
can be used to build, for instance, the perturbative filtered spectral function.
In Fig.~\ref{fig:BG,rhohat}, we compare the filtered spectral function
obtained from the continuum $G_\mathrm{T}/T^3$ data to the filtered spectral function
of the weak-coupling QCD regime as well as of the 
SYM theory, using the same resolution functions.
The chosen rescaling function was
\beq
f(\om)= \frac{\om^2}{\tanh(\om/(2T))}.
\eeq

\begin{figure}[t!]
	\includegraphics[scale=0.64]{{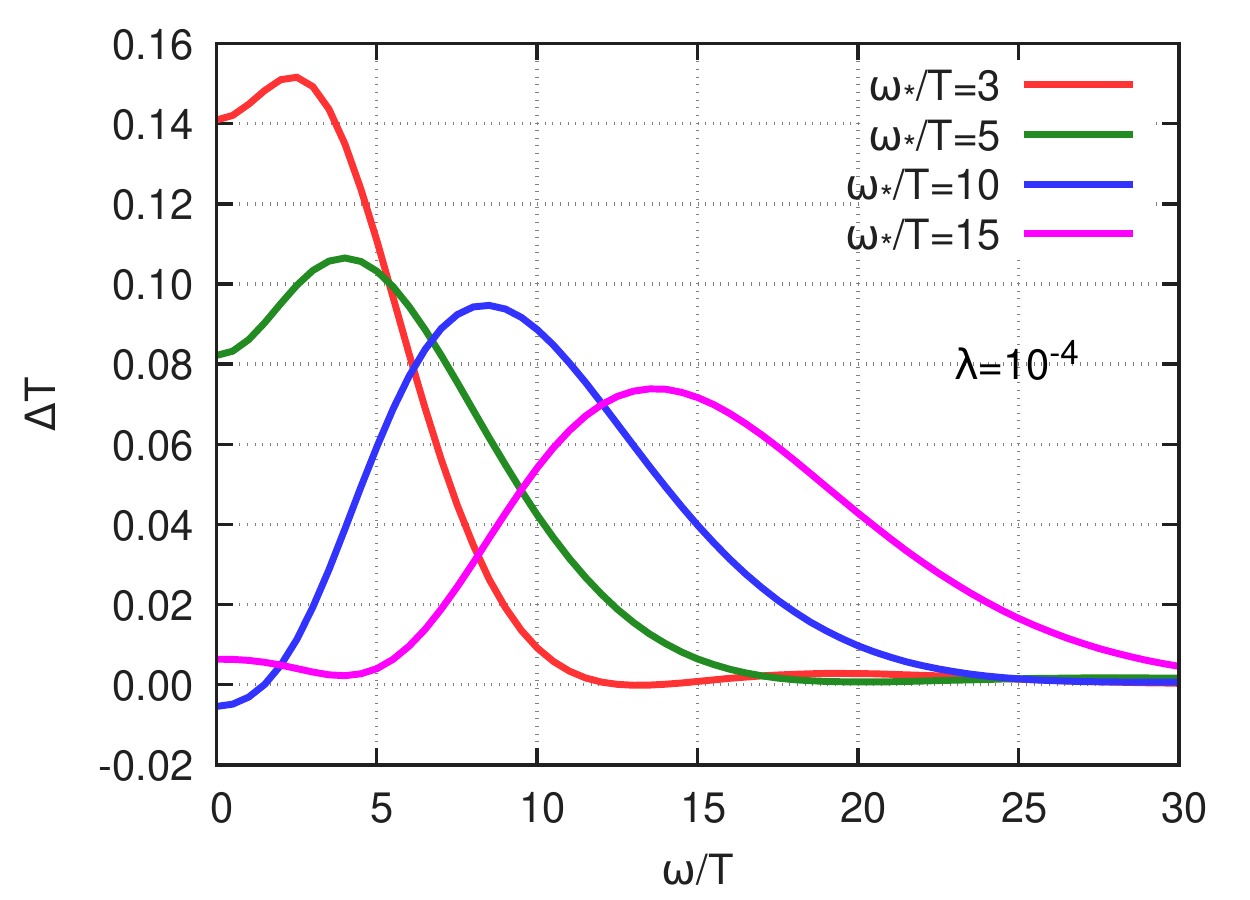}}
\caption
{
	\label{fig:BG,TDelta}
	Resolution functions for $k/T \approx 3.14$, $\lambda=10^{-4}$ for different
	values of $\om_*/T$.
}
\end{figure}

As Fig.~\ref{fig:BG,rhohat} shows, the filtered, rescaled spectral function
obtained using the lattice data is somewhat below the perturbative one,
especially for small momenta ($k/T \le 3.14$).
At small frequencies, below $\om/T \sim 5$, the filtered spectral function
is between the weak-coupling and the $\mathcal{N}=4$ SYM 
theory filtered spectral function.
Using all nine data points for the correlator, we found that the
condition number of ${\bf A}^{\rm{reg}}$ gets more or less tractable (smaller than $\sim 10^7$)
when $\lambda$ is around $10^{-4}$ or smaller.
The elements of ${\bf A}^{\rm{reg}}$ are still dominated in this case by the first
term of Eq.~(\ref{eq:Aijreg}), which are several orders of magnitude larger than the 
elements of the covariance matrix.
When $\lambda \gtrsim 10^{-2}$, the condition number is above $5 \times 10^8$ and
$\hat{\rho}(\om)$ tends to have unphysical wiggles.
By omitting e.g.\ the first data point from the correlator, one can use larger
values of $\lambda$, but one less coefficient.
The results, nevertheless, do not change significantly.

The resolution functions, $\Delta(\om,\om^\prime)$, are quite wide and they
blur the fine details of the spectral function.
They are almost identical for different momenta when fixing the value of $\lambda$.
Their dependence on $\lambda$ is also very mild, but increases when going 
to a higher target $\om$.
Some examples can be seen in Fig.~\ref{fig:BG,TDelta}.
We obtained similar results to those discussed above using the covariance matrix of $G_\mathrm{T}/G_{00}$ instead.

In summary, we have found that the filtered spectral functions derived
from the continuum-extrapolated lattice data are close to the NLO+LPM
spectral function filtered with the same resolution function, and that the
agreement increases both with increasing $k$ and increasing $\om$, up
to $\om\approx 10T$. Since beyond that point the resolution function
extends to very high frequencies, residual cutoff effects in the
lattice results could explain the differences observed with the
NLO+LPM curves in that regime.  At $k=\pi T$, the non-perturbative
filtered spectral function undershoots the perturbative prediction, a
point already noted in~\cite{Ghiglieri:2016tvj} for the $2\rho_{\rm T}+\rho_{\rm L}$
channel relevant for the dilepton rate.
Finally, we remark that the value of $[\hat\rho(\om,k)/f(\om)]_{\om=k}$ found here is very much in the ballpark
of the values obtained for $[\rho(\om,k)/f(\om)]_{\om=k}$ with various fit ans\"atze for the spectral function
(see Fig.~\ref{fig:pipoly,rho} below), even though the resolution function is rather broad compared to $k$.

\subsection{The transverse-channel spectral function from fits to the Euclidean correlators}
\label{sec:fitmethod}
We now turn to turn to our direct attempt at determining
the spectral function by fitting various ans\"atze to the transverse correlator.
When specifying possible ans\"atze describing the transverse spectral
function, we have taken into account the facts that it has to be odd
in $\om$, i.e. $\rho(-\om) = -\rho(\om)$, as well as 
positive for $\om>0$, as shown in Ref.~\cite{Ce:2020tmx}.
While the ans\"atze were chosen to be odd by construction, models were only
excluded a posteriori if they violated positivity with a 68\% confidence level.

The high-frequency behavior of the spectral function is dictated 
by perturbation theory, therefore we assumed the following form
for the transverse spectral function
\beq
	\rho(\om) = \rho_{\sfit}(\om) ( 1-\Theta(\om, \om_0, \Delta) )
			  + \rho_{\spert}(\om) \Theta(\om, \om_0, \Delta),
	\label{eq:rhoT,fit}
\eeq
where $\rho_{\spert}$ is the NLO prediction for the transverse spectral 
function complemented with the LPM contribution near the light-cone,
\beq
	\rho_{\spert}(\om) = \rho_{\mathrm{NLO}}(\om) + \rho_{\mathrm{LPM}}(\om),
\eeq
and
\beq
\Theta(\om, \om_0, \Delta) = (1 + \tanh[(\om-\om_0)/\Delta]) / 2
\eeq
is a smooth step function which controls how fast the 
perturbative contribution falls off around $\om_0$ as $\om$ is lowered.
Using the decomposition of Eq.~(\ref{eq:rhoT,fit}), we ensure
that the perturbative part gives the dominant contribution above 
the chosen value of $\om_0$, which we call the matching frequency.
Moreover, the transition from the perturbative regime --- assumed to be
valid in the ultraviolet --- can be realized in a smooth way without
constraining any of the coefficients of the fit function.

For the fit function, $\rho_{\sfit}(\om)$, we considered the following 
two possible ans\"atze:
\beq
\frac{\rho_{\mathrm{fit,1}}(\om)}{T^2} = \sum_{n=0}^{N_\mathrm{p}-1} A_n \left(\frac{\om}{\om_0}\right)^{1+2n},
	\label{eq:fit,poly}
\eeq
where $N_\mathrm{p}$ denotes the number of fit parameters, and
\beq
\frac{\rho_{\mathrm{fit,2}}(\om)}{T^2} = \,
\begin{dcases}
	\,A_0 \,\frac{\om}{\om_0} + A_1 \left(\frac{\om}{\om_0}\right)^3, & \text{if\,\,} \om \le k, \\
	\,B_0 \,\frac{\om}{\om_0} + B_1 \left(\frac{\om}{\om_0}\right)^3, & \text{if\,\,} \om > k,
	\label{eq:fit,pipoly}
\end{dcases}
\eeq
where the free parameters have been chosen to be $A_0, B_0, B_1$, and 
$A_1$ has been fixed imposing continuity, $A_1 = B_1 + (B_0 - A_0) \om_0^2/k^2$.
In the latter case, we also carried out fits by setting $B_0$ to zero,
i.e.\ having only two fit parameters.
The spectral function of Eq.~\ref{eq:rhoT,fit} including $\rho_{\mathrm{fit,1}}(\om)$
or $\rho_{\mathrm{fit,2}}(\om)$ is referred to in the following as the polynomial
or the piecewise polynomial ansatz, respectively.

\begin{figure}[t!]
	\includegraphics[scale=0.64]{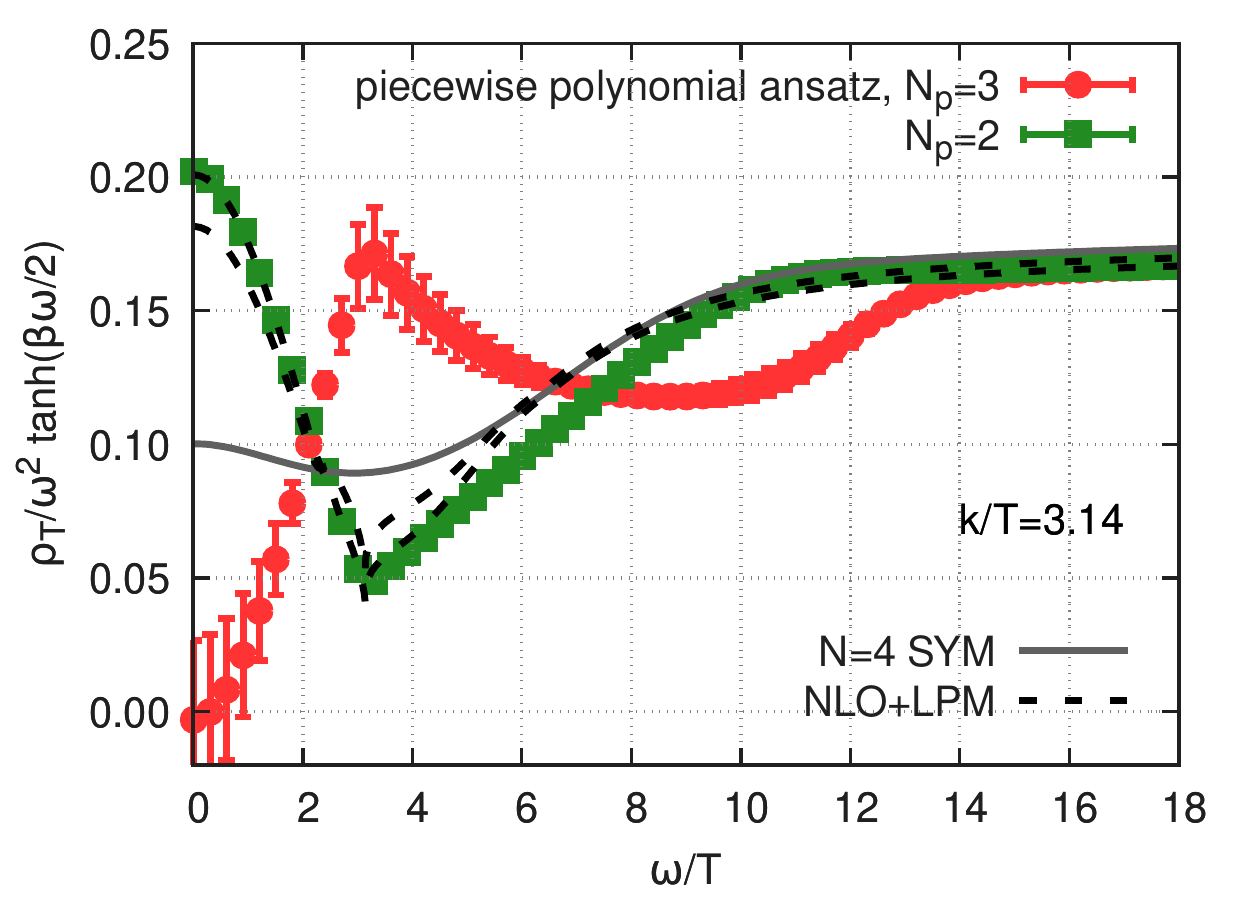}
	\includegraphics[scale=0.64]{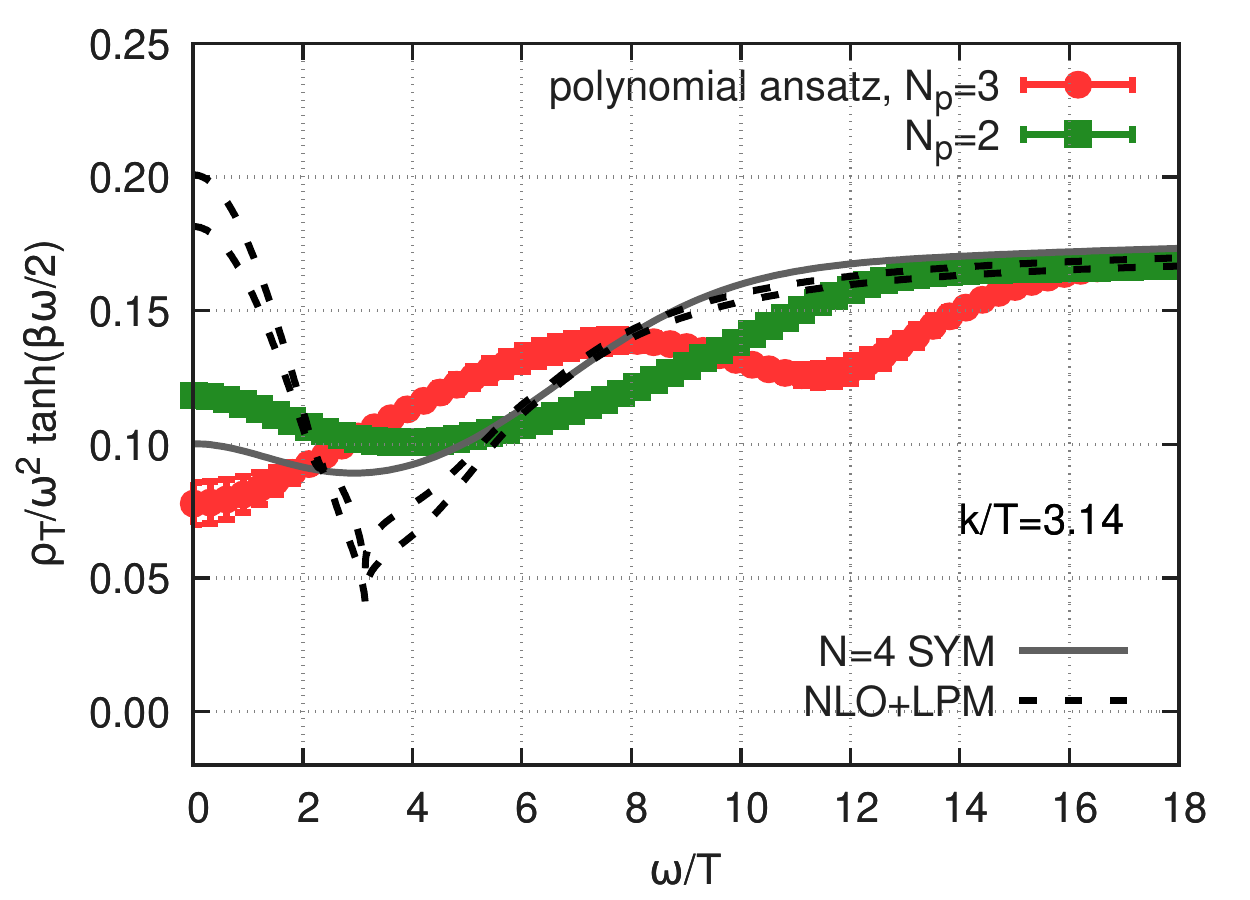}
\caption
{
	{\bf Left}: Representative fit results using the piecewise polynomial ansatz
	with two ($N_p=2$) or three fit parameters ($N_p=3$) at $k/T=\pi$.
	For comparison, the transverse channel spectral function in the
	weak-coupling NLO+LPM theory, as well as in the strongly coupled
	SYM theory are also included.
	{\bf Right}: Representative fit results using the polynomial ansatz.
	\label{fig:pipoly,rho}
}
\end{figure}

These ans\"atze are motivated by their ability to describe the
spectral function of the strongly coupled $\mathcal{N}=4$ SYM theory
as well as that of the NLO+LPM resummed perturbation
theory, to a satisfactory level.  The functional form of
Eq.~(\ref{eq:fit,poly}) is more suitable for the spectral function
obtained with the AdS/CFT approach in the $\mathcal{N}=4$ SYM theory,
whereas the built-in non-differentiability at the light-cone in the
piecewise polynomial ansatz of Eq.~(\ref{eq:fit,pipoly}) is more apt
at describing the cusp present in the NLO+LPM result. More details
can be found about the expressivity of these ans\"atze in Appendix
\ref{app:mock}, in which we present the outcome of mock data analyses
and to which we return in the next subsection.

After inserting $\rho(\om)$ from Eq.~(\ref{eq:rhoT,fit}) into
Eq.~(\ref{eq:sdecomp1}), we solved the correlated
$\chi^2$-minimization problem to determine the unknown coefficients.
We show some representative fit results with good $\chi^2$ values in Fig.~\ref{fig:pipoly,rho}
for the momentum $k =\pi T$. We explored many variations in the fit
procedure and took into account the statistical and systematic error
of the continuum-extrapolated Euclidean correlators. More details on
this and the method of estimating systematic errors on the
effective diffusion coefficient, $D_{\rm eff}(k)$,
can be found in Appendix~\ref{app:techasp}.
Our result for $D_{\rm eff}(k)$ extracted using the polynomial ansatz
is displayed on the right panel of Fig.~\ref{fig:TDeff,ncd}.
The line within the band represents the median of the distribution of results obtained,
while the width of the band indicates the position of the 16th and 84th percentile.

The piecewise polynomial fit ansatz turned out to yield
a sizeable spread of results for the effective diffusion coefficient.
This spread comes from  the results for $D_{\mathrm{eff}}$ actually falling into two well separated
intervals. Fit results with $N_\mathrm{p}=2$ tend to lead to results in the lower interval,
while the results obtained with $N_\mathrm{p}=3$ populate both intervals.
To illustrate the point, on the left panel of Fig.~\ref{fig:TDeff,ncd} we display the results
for the effective diffusion coefficient $D_{\rm eff}(k)$
obtained with $N_p=2$ and $N_p=3$ separately as two coloured bands.
This doubly-peaked distribution of results for $D_{\mathrm{eff}}$ is associated with the
behaviour of the spectral function around lightlike kinematics.
When using this ansatz to fit the correlator,
we obtained spectral functions possessing either a minimum or a spike-like maximum
at the light-cone frequency. Representative fit results are shown on the left panel of Fig.~\ref{fig:pipoly,rho}.
Since neither the NLO+LPM weak-coupling nor the SYM spectral
functions have a maximum at $\om = k$, we also investigated the fit results
obtained by excluding the fits satisfying
\beq
    \frac{\sd \rho}{\sd \om}\Big|_{\om = k-\epsilon} - \frac{\sd \rho}{\sd \om}\Big|_{\om = k+\epsilon} > 0
	\label{eq:no-peak_cond}
\eeq
at least at one standard deviation.
With this qualitative theoretical prejudice in place, the fit results are significantly more constraining.
The $D_{\rm eff}$ values obtained this way are displayed on the right panel of Fig.~\ref{fig:TDeff,ncd}, where
they are denoted as ``no-peak'' solutions.
Excluding fits possessing the feature Eq.~(\ref{eq:no-peak_cond}) is also affirmed
by the upper limit of the results obtained by analysing the spectral function
in the $\mathrm T-\mathrm L$ channel~\cite{Brandt:2019shg,Ce:2020tmx}.
As can be seen by comparing the two panels of Fig.~\ref{fig:TDeff,ncd},
a large portion of the solutions with a peak at lightlike kinematics 
can be excluded by our previous analysis of the $\mathrm T-\mathrm L$ channel.
We note that the latter analysis has been carried out using the same
ensembles that we employ in the present study.
For comparison, Fig.~\ref{fig:TDeff,ncd} also displays the weak-coupling results
(dashed lines) obtained directly for the photon rate in Ref.~\cite{Arnold:2001ms},
as well as the strongly coupled SYM theory results (solid grey line).

\begin{figure}[t!]
	\includegraphics[scale=0.64]{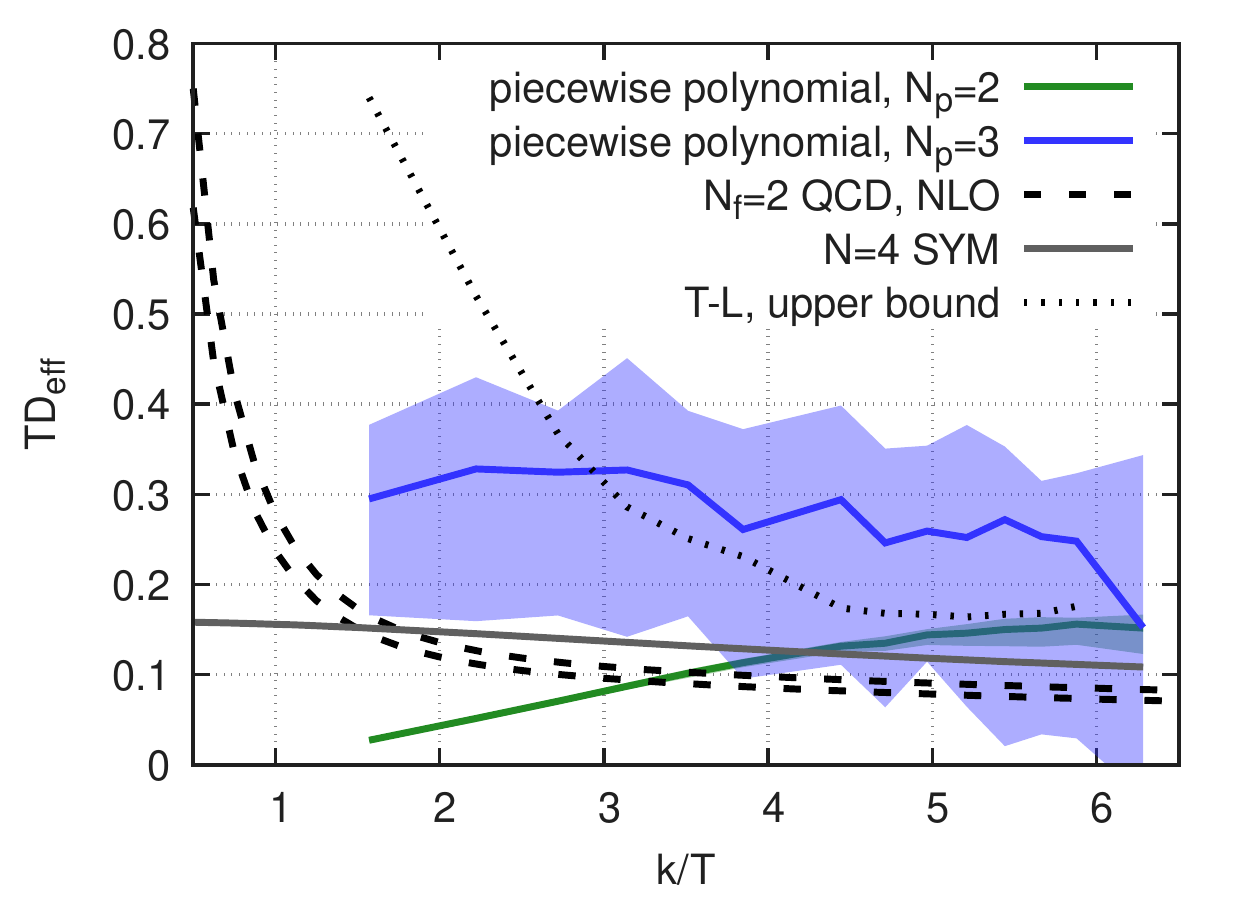}
	\includegraphics[scale=0.64]{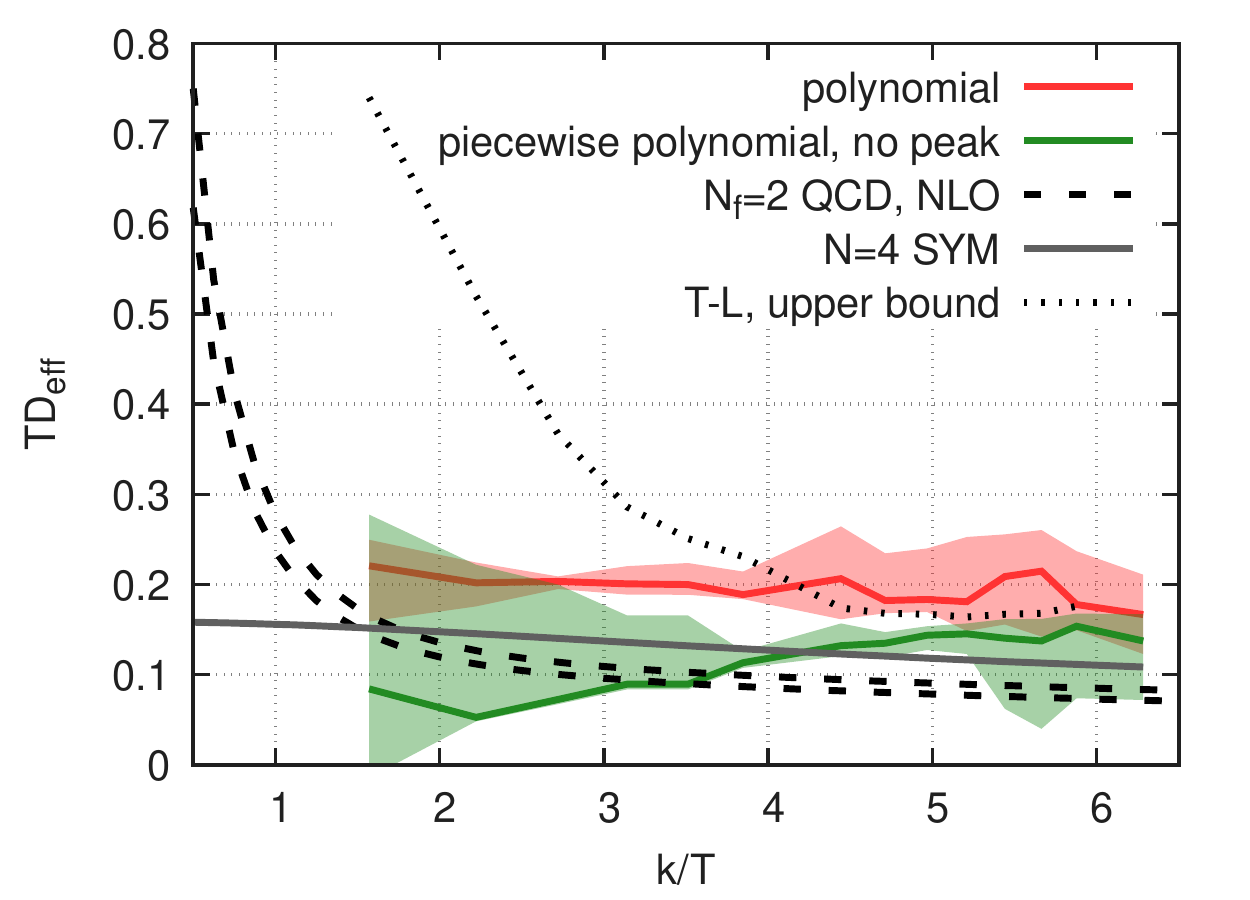}
\caption
{
	Results for the effective diffusion coefficient, $T D_{\mathrm{eff}}$,
	defined in Eq.~(\ref{eq:TDeff}).
	{\bf Left}: Separate analyses of fit results obtained
	with $N_\mathrm{p}=2$ and $N_\mathrm{p}=3$ using the piecewise polynomial ansatz (Eq.~(\ref{eq:fit,pipoly})).
	{\bf Right}: Imposing the constraint of Eq.~(\ref{eq:no-peak_cond})
	at one standard deviation to the fit results obtained with the
	piecewise polynomial ansatz. The results obtained by using the polynomial ansatz (Eq.~(\ref{eq:fit,poly})) are displayed in red.
	Analytical results from perturbative QCD~\cite{Arnold:2001ms} using $\alpha_s\simeq0.25$
    or 0.31 (dashed lines) and from the strongly coupled $\mathcal{N}=4$ SYM theory (grey line)~\cite{Caron-Huot:2006pee}
	are also included as well as an upper bound
    obtained from the analysis of the $\mathrm T-\mathrm L$ channel (dotted line)~\cite{Ce:2020tmx}.
    We use the value $\chi_\mathrm{s}/T^2=0.88(2)$ to obtain $T D_{\rm{eff}}$ from the lattice data,
	but the free susceptibility to obtain $T D_{\rm{eff}}$ from the $N_{\rm f}=2$ weak-coupling photon rate.
	\label{fig:TDeff,ncd}
}
\end{figure}
\subsection{Final result for the photon emissivity extracted from the fits}
\label{sec:finalres}
In order to arrive at our final estimate for the photon emissivity, we
need to judge the reliability of our fit ans\"atze in extracting the
quantity $D_{\rm eff}$. For this we return to the analysis of mock
data, presented in detail in Appendix~\ref{app:mock}. There, we found
that the fit ansatz functions tend to overestimate the photon rate in
the investigated models, irrespective of whether $G_\mathrm{T}/G_{00}$ or
$G_\mathrm{T}/T^3$ is used. We therefore first derive an upper bound
for $D_{\rm eff}$, and hence for the photon emissivity.
By fitting a linear ansatz $a \times k/T + b$ to $T D_{\mathrm{eff}}$
obtained from the polynomial ansatz
in the $k/T$ range $[\pi/2, 2\pi]$ and then plugging the effective diffusion
coefficient, $T D_{\mathrm{eff}}$, into the formula giving the thermal photon
emission rate per unit volume, Eq.~(\ref{eq:photonrate}), we arrive at the following upper bound,
\begin{equation}
	\frac{1}{T} \frac{\sd \Gamma_\gamma(k)}{\sd^3 k} \hspace{-0.1cm} \leq
        \frac{\alpha_\mathrm{em}}{\pi^2} \, \frac{\chi_\mathrm{s}}{T^2} \, \frac{C_\mathrm{em}}{\mathrm e^{k/T}-1} \text{\small{$\times$}}
        \Big( 0.22(1) -0.0062(25) \frac{k}{T}  \Big).
	\label{eq:photonrate,upper}
\end{equation}
In Eq.~(\ref{eq:photonrate,upper}), $C_\mathrm{em}$ denotes the charge factor
equal to $2/3$ in the $N_\mathrm{f}=3$ theory.
We recall that the static susceptibility has been determined in
Sec.~\ref{sec:contlim}, $\chi_\mathrm{s}/T^2 = 0.88(2)$ at $T\simeq 250$~MeV.

\begin{figure}[t!]
	\includegraphics[scale=0.64]{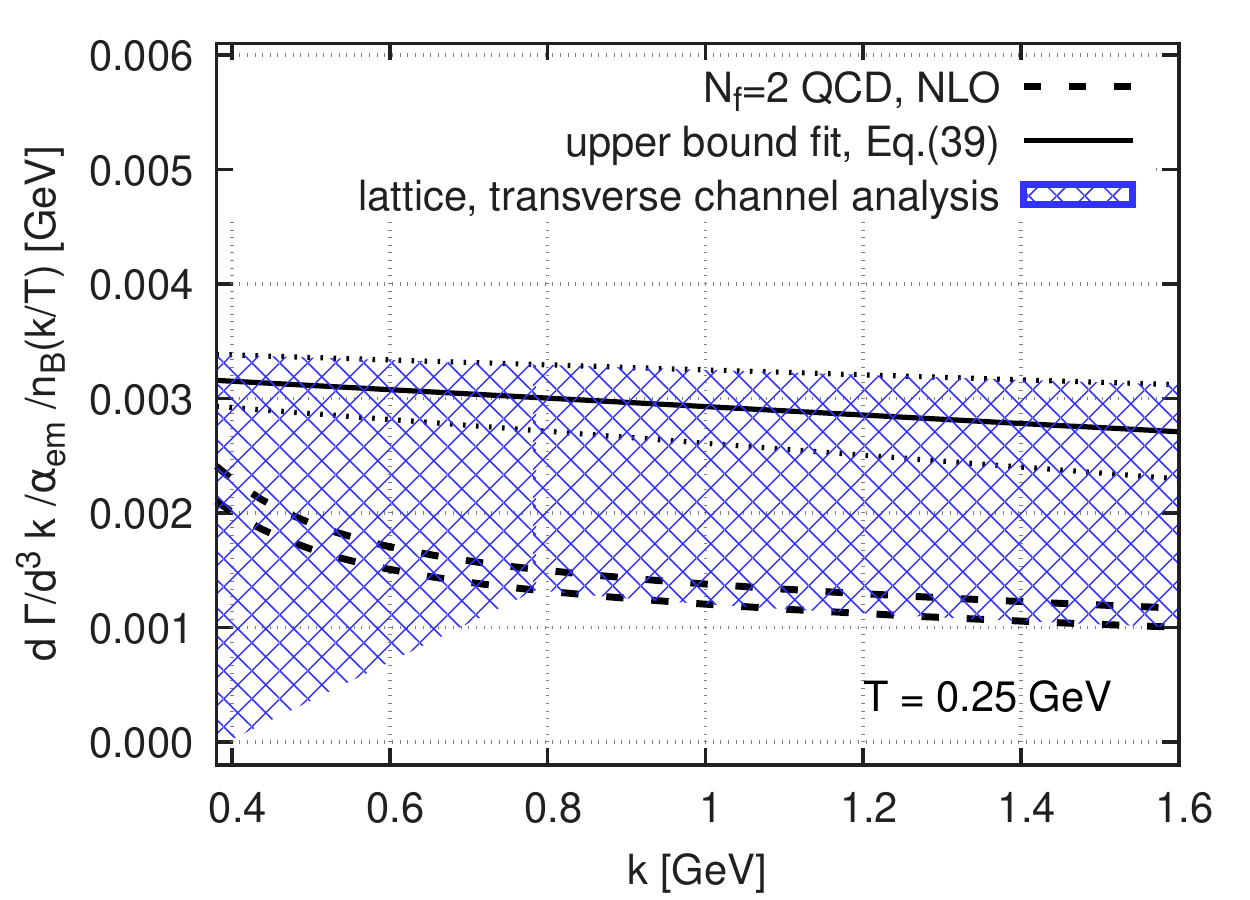}
\caption
{
	The photon rate per unit volume of the QGP at $T\simeq250$ MeV (hatched band).
  The Bose factor  $n_\mathrm{B}=1/(\mathrm e^{k/T}-1)$ has been divided out.
	The upper bound parametrization derived from the $D_{\rm eff}$ results of the polynomial ansatz
	is shown as a solid line.
	The full leading-order result~\cite{Arnold:2001ms} obtained by setting $\alpha_\mathrm{s}\simeq0.25$ 
	or 0.31 are shown for comparison.
	Including corrections from interactions with soft gluons, one can
    get a 20\% increase at $\alpha_\mathrm{s}\simeq 0.3$ for the perturbative result~\cite{Ghiglieri:2013gia}.
	\label{fig:photon_rate}
}
\end{figure}

As far as a lower bound on $D_{\rm eff}$ is concerned, 
we distinguish two regimes, $k>\pi T$ and $k\leq \pi T$.
In the former case, using the NLO+LPM correlator as input,
the central value of the output for the effective
diffusion constant is usually at least $50 - 80 \%$ larger than the
true value, and our error estimate typically does not cover the true value
in the range $3.5\leq k/T \leq 5.5$ (see the right panel of Fig.~\ref{fig:mock,nlo3,s=1,ncd}).
In the case of the $\mathcal{N}=4$ SYM theory mock data, the results
from the two ans\"atze bracket the true result for intermediate and
smaller momenta $k/T \lesssim 4.5$, but our ans\"atze overestimate
$TD_{\mathrm{eff}}$ by $5 - 25\%$ at larger momenta.
Therefore, for $k>\pi T$ we extend the lower bound down to
the weak-coupling prediction in Fig.~\ref{fig:TDeff,ncd} (right panel),
which we parametrize in Eq.\ (\ref{eq:photonrate,lower}).

For the mock data tests performed in the momentum range $k\leq \pi T$ on the other hand,
the resulting error band always covers the true value for at least one of the two ans\"atze that we use.
As the right panel of Fig.~\ref{fig:TDeff,ncd} shows, the lower bound is
driven by the results of the piecewise polynomial ansatz, which decreases
until reaching vanishing $D_{\rm eff}$ values at our smallest momentum.
A parametrization of the lower bound can be given by a linear function
in $k/T$ that connects zero at $k = \pi T/2$ and the NLO weak-coupling
result at $k= \pi T$. Thus we parametrize our lower bound according to
\begin{equation}
	\frac{1}{T} \frac{\sd \Gamma_\gamma(k)}{\sd^3 k} \hspace{-0.1cm} \geq
        \frac{\alpha_\mathrm{em}}{\pi^2} \, \frac{\chi_\mathrm{s}}{T^2} \, \frac{C_\mathrm{em}}{\mathrm e^{k/T}-1} \text{\small{$\times$}}
	\left\{ \begin{array}{ll}        \big( -9.43 + 6.00\, \frac{k}{T} \big) \times 10^{-2},  & \pi T/2 \leq k \leq \pi T;
          \phantom{\Big|}\\
		\big( 4.86 + 14.33 \, \frac{T}{k} \big) \times 10^{-2}, & \pi T < k \leq 2\pi T.
          \end{array}\right.
	\label{eq:photonrate,lower}
\end{equation}
From the observations made above, it is clear that this lower bound is influenced by the theory prejudice that the spectral function
at lightlike kinematics does not `dive' down to even smaller values than the weak-coupling results~\cite{Arnold:2001ms} predict
for realistic values of $\alpha_s$; see Fig.~\ref{fig:pipoly,rho}.

We remark that the curves and bands displayed in Fig.~\ref{fig:TDeff,ncd} have a
precise statistical meaning based on percentiles ($16^{\rm th}$, $50^{\rm th}$, $84^{\rm th}$) of the distribution
of results obtained by applying a set of procedural variations.
The band defined by the $k$-dependent bounds of Eqs.~(\ref{eq:photonrate,upper},\ref{eq:photonrate,lower}) summarizes the results,
giving equal weight to both fit ans\"atze that we have used; it is displayed in Fig.~\ref{fig:photon_rate}.

\section{Conclusion}
\label{sec:concl}
In this study, we have presented the first investigation of the transverse channel
Euclidean correlator at finite momenta, using 
$N_{\mathrm{f}}=2$ $\mathrm{O}(a)$-improved dynamical Wilson fermions
at around $T\simeq 250\,\mathrm{MeV}$ corresponding to $1.2T_\mathrm{c}$.
We carried out a simultaneous continuum extrapolation using three 
discretizations of the correlators of the isovector vector currents.
We used four ensembles with lattice spacings in the range $a\simeq 0.033\,$--$\,0.066$\;fm.
We compared the filtered spectral functions obtained from the continuum
extrapolated transverse correlator via the Backus-Gilbert method
to the spectral function of perturbation theory at next-to-leading order and the one obtained in
the strongly coupled $\mathcal{N}=4$ supersymmetric Yang-Mills theory.
The small-frequency lattice results lie between the filtered spectral
functions obtained in these two theories.
In order to determine the thermal photon emission rate, we fitted the
correlators using polynomial and piecewise polynomial fit ansätze for 
the underlying spectral function.
We validated the expressivity of these ansätze by performing mock tests
using the spectral functions obtained in perturbation theory as well as
in the $\mathcal{N}=4$ super Yang-Mills theory.
We compared our results for the effective diffusion coefficient
to the results obtained in these theories as shown in Fig.~\ref{fig:TDeff,ncd},
and also to the results obtained by analysing the difference of the
transverse and the longitudinal channels, presented earlier in Ref.~\cite{Ce:2020tmx}.
The ranges obtained are compatible with the weak-coupling as well as with the AdS/CFT results.
Our final result for the photon emissivity is displayed in Fig.~\ref{fig:photon_rate} as a hatched band,
for which we provide a parametrization in Eqs.~(\ref{eq:photonrate,upper},\ref{eq:photonrate,lower}).
For momenta below 1\,GeV, the upper edge of the band is more constraining than
the estimate derived from the analysis of the difference of the transverse
and longitudinal channels. We obtain our strongest constraint on the photon emissivity around $k=\pi T \simeq 0.8\,$GeV,
\begin{equation}
  \frac{\sd\Gamma_\gamma}{\sd^3k} =
  \frac{\alpha_{\rm em}}{e^{k/T}-1}\, (2.2 \pm 0.8 ) \times 10^{-3}\,{\rm GeV}.
\end{equation}

In the future, we plan to (linearly) combine the present analysis with our
previous one~\cite{Ce:2020tmx} to provide estimates of the dilepton
rate at invariant masses $M_{\ell^+\ell^-}^2 = \omega^2-k^2>0$.  We
are also investigating a qualitatively different approach to the
photon rate by computing zero-virtuality correlators on the
lattice~\cite{Meyer:2018xpt,Ce:2021nvy}. Whilst numerically
challenging, this approach allows one to avoid confronting the inverse
problem.


\acknowledgments
We thank Arianna Toniato for discussions and valuable input in the early stages of this project.
This work was supported by the European Research Council (ERC) under the European 
Union’s Horizon 2020 research and innovation program through Grant Agreement
No.\ 771971-SIMDAMA, as well as by the Deutsche Forschungsgemeinschaft 
(DFG, German Research Foundation) through the Cluster of Excellence “Precision Physics,
Fundamental Interactions and Structure of Matter” (PRISMA+ EXC 2118/1) funded by
the DFG within the German Excellence strategy (Project ID 39083149).
M.C.\ was supported by the European Union's Horizon
2020 research and innovation program under the Marie Sk\l{}odowska-Curie Grant Agreement 
No.\ 843134-multiQCD.
T.H. is supported by UK STFC CG ST/P000630/1.
The generation of gauge configurations as well as the computation of correlators was 
performed on the Clover and Himster2 platforms at Helmholtz-Institut Mainz and on Mogon II 
at Johannes Gutenberg University Mainz.
We have also benefitted from computing resources at Forschungszentrum J\"ulich allocated 
under NIC project HMZ21.
We made use of the following libraries: GNU Scientific Library~\cite{GSL:2019},
GNU MPFR~\cite{10.1145/1236463.1236468} and GNU MP~\cite{GMP:web}.

\appendix

\section{Technical aspects of the fits for the spectral function \label{app:techasp}}

In this appendix, we provide some details on the procedure we followed
in Sec.~\ref{sec:fitmethod} to obtain spectral functions via fits to the Euclidean correlators.
After inserting $\rho(\om)$ from Eq.~(\ref{eq:rhoT,fit}) into 
Eq.~(\ref{eq:sdecomp1}), we solved the correlated $\chi^2$-minimization 
problem to determine the unknown coefficients.
When doing so, we regularized the covariance matrix by multiplying 
the off-diagonal elements by 0.95.
To estimate the statistical error, the minimization problem has been 
solved for every jackknife sample of the continuum correlator by using 
the covariance matrix obtained from the full data.
To quantify the systematic uncertainty of the reconstruction 
coming from the systematic uncertainty of the continuum limit correlator
values, we used the continuum extrapolation fit results for the 16th, 50th
and 84th percentile of the AIC-weighted histogram obtained for each $\tau T$ value.
However, we did not take into account all possible combinations of these
histogram representatives, but chose random subsets containing 20 -- 100
combinations for each momentum. 

In order to estimate the systematic error from making 
parameter choices for the perturbative results or changing various
parameters when fitting, we performed several fits using a set of 
plausible variants.
Regarding the perturbative input for our analysis, the NLO+LPM result 
depends on the coupling constant.
Using the four-loop formulae of Ref.~\cite{Deur:2016tte}, we determined
the coupling constant by setting the renormalization scale either
to $\mu = 2\pi T$ or to $\mu = 3\pi T$.
The coupling constants corresponding to these choices are $\alpha_\mathrm{s}\simeq0.31$
and $\alpha_\mathrm{s}\simeq0.25$, respectively.
We took $\Lambda_{\overline{\rm MS}}^{(N_\mathrm{f}=2)}$
from the FLAG report~\cite{Aoki:2019cca}.

\begin{table}[b]
\begin{tabular}{ll}
\hline
\hline
{\bf \, source of systematic error} & {\,\,\,\,\bf variations} \\ \hline
\#(correlator data pts) & \,\,\,\,9, 8, 7, 6\\
\#(fit parameters)  & \,\,\,\,2, 3 \\
soft: $\Delta/T$    & \,\,\,\,1.6, 2.0, 2.4 \\
$\om_0/T$			& \,\,\,\,10, 12 \\
ren. scale 			& \,\,\,\,$(2, 3)\times \pi T$ \\
\hline
\hline
\end{tabular}
\caption{
\label{tab:fitvars}
Summary of the systematic variations we employed in the fit approach.
Using nine (six) data points correspond to fitting from $\tau T \approx$ 0.167 (0.292).
}
\end{table}

We applied two values for the matching frequency, $\om_0 = 10 T$ and 
$\om_0 = 12 T$, which correspond to 2.5 GeV and 3 GeV, respectively.
For the parameter $\Delta$ in $\Theta(\om,\om_0,\Delta)$, which governs 
the falloff of the NLO+LPM contribution towards the infrared, we set 
$\Delta = 2 T$, allowing also for a 20\% variation.
As a further systematic variation, we adjusted the fit ranges in
$\tau T$ to include more or fewer data points in the fits.
The variations we applied for the systematic error estimation
are summarized in Table~\ref{tab:fitvars}.
Due to these variations, we collected 750--2000 fits with $p$-values larger
than 0.05 for each momentum when using the polynomial ansatz, and around 15--30\%
of these fits had $p$-values larger than 0.5.
When using the piecewise polynomial ansatz, we also obtained a lot of
fits with good $\chi^2$ (and $p$-values). The fraction of the number of fits
with $p$-values greater than 0.5 was around 15--30\% in that case as well.
We built an AIC-weighted histogram from the fit results that
we used to estimate the systematic errors~\cite{Akaike:1973abc,Borsanyi:2020mff}.

\section{Mock analyses}
\label{app:mock}

We performed mock analyses to see whether the fit ansätze are expressive
enough to reproduce the transverse channel spectral function as well
as to investigate the reliability of the Backus-Gilbert method.
For these tests, we used two models: 
\bit
	\item[(i)] the NLO weak-coupling spectral function complemented
with the LPM contribution near the light-cone and 
	\item[(ii)] the spectral function of the strongly coupled $\mathcal{N}=4$ super 
Yang-Mills (SYM) theory.
\eit

We generated mock correlators using an appropriately rescaled covariance
matrix of the lattice covariance matrix, which was obtained with the help
of the continuum extrapolated correlator and the jackknife samples.

The procedure we followed consists of the following steps:
\ben
	\item calculation of the ratio, $r(\tau)$, of the model and the continuum
		lattice correlator, $r(\tau) := G_\mathrm{model}(\tau)/G(\tau)$;
	\item rescaling the covariance matrix of $G$ in the continuum:
		$\Cov_{\mathrm{resc},\,\tau \tau^\prime} := \Cov_{\tau \tau^\prime} \, r(\tau) r(\tau^\prime)$;
	\item generating multivariate Gaussian variables, $G_\mathrm{mock}(\tau)$, using the
		rescaled covariance matrix, $\Cov_\mathrm{resc}$, as well as $G_\mathrm{model}(\tau)$;
	\item determination of the ratio $\bar{r}(\tau)$ of the mock correlator,
		$G_\mathrm{mock}(\tau)$, and the lattice continuum correlator;
	\item rescaling all jackknife samples using $\bar{r}(\tau)$
		and calculating the covariance matrix of the mock correlator ($\Cov_\mathrm{mock}$)
		using these rescaled values.
\een

We used then $G_\mathrm{mock}(\tau)$ and $\Cov_\mathrm{mock}$ in the fit analysis.
The mock jackknife samples have been used to estimate the statistical error
in the mock analysis.
The relative errors of the mock correlator are roughly the same as
the relative errors of the continuum extrapolated correlator due
to the rescaling.
We found that applying only step 1 and step 2 of the procedure above,
complemented with a simple rescaling of the jackknife samples with $r(\tau)$
results in the overestimation of errors.

In addition to using the same relative errors as the continuum lattice data,
we also investigated the effect of reducing the errors on the
mock correlator and therefore included slight modifications to the above
procedure.
Introducing the error reduction factor, $s>1$, we used $r(\tau) / s$
in step 2, and instead of a rescaling, we determined the $j$th jackknife
mock correlator value as
$\bar{G}_\mathrm{mock}^{(j)}(\tau) = \bar{G}_\mathrm{mock}(\tau)
    + \left( r(\tau) \bar{G}^{(j)}(\tau) - \bar{G}_\mathrm{mock}(\tau) \right)/s$
in step 5.

Since we extrapolated $\widebar{G} = G_\mathrm{T}/G_{00}$ to the continuum
(see Sec.~\ref{sec:contlim}), we also carried out the mock analysis using
this observable.
When doing so, $G$, $G_\mathrm{mock}$ and $G_\mathrm{model}$ should be replaced by
$\widebar{G}$, $\widebar{G}_\mathrm{mock}$ and $\widebar{G}_\mathrm{model}$,
respectively, in the above procedure, and we refer to this ratio of
the chosen model when we have e.g. $\widebar{G}_\mathrm{model}$.
Since the covariance matrices are quite different for $G_\mathrm{T}/T^3$ and for
$G_\mathrm{T}/G_{00}$, the mock analyses starting from the former or the latter
result in different outcomes.

\begin{figure}[t!]
	\includegraphics[scale=0.60]{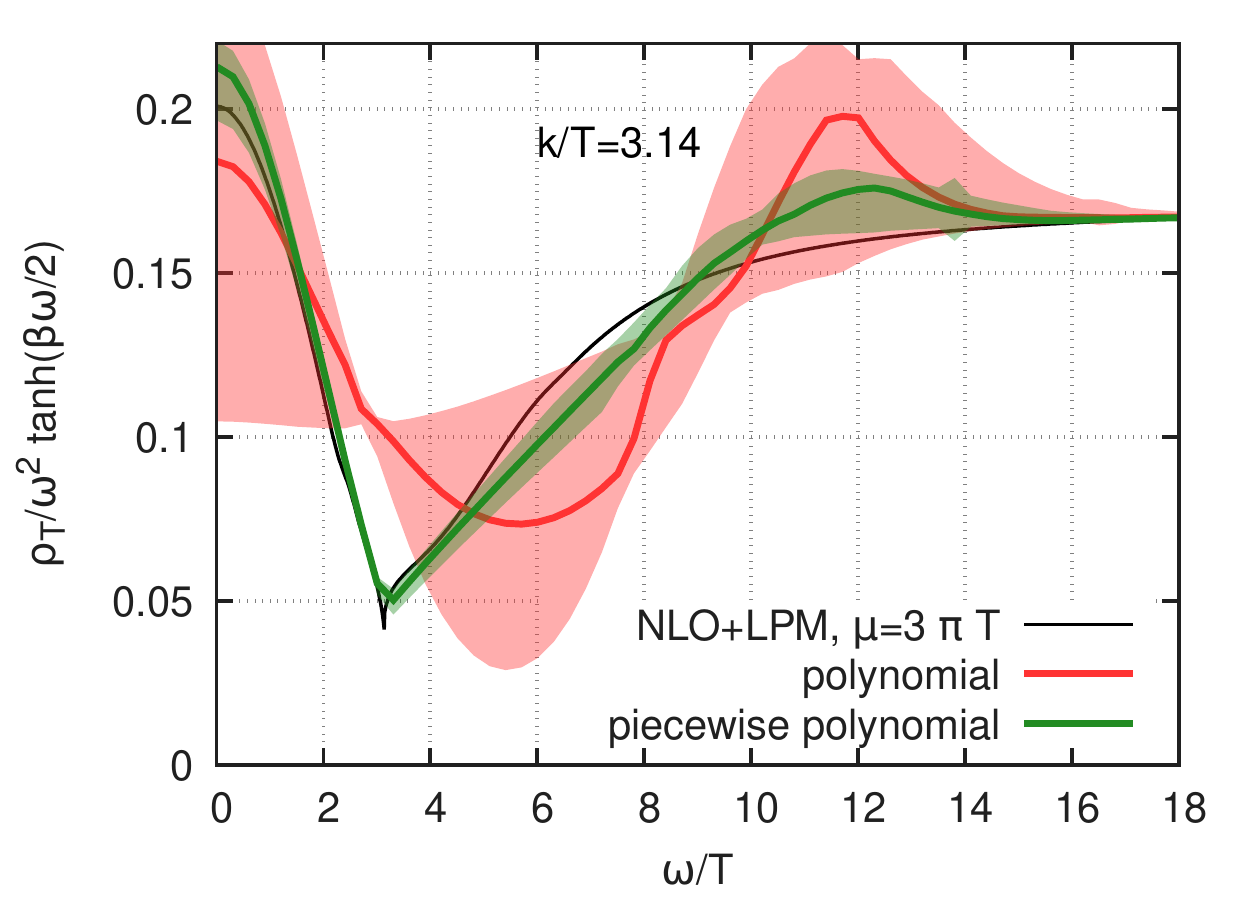}
	\includegraphics[scale=0.60]{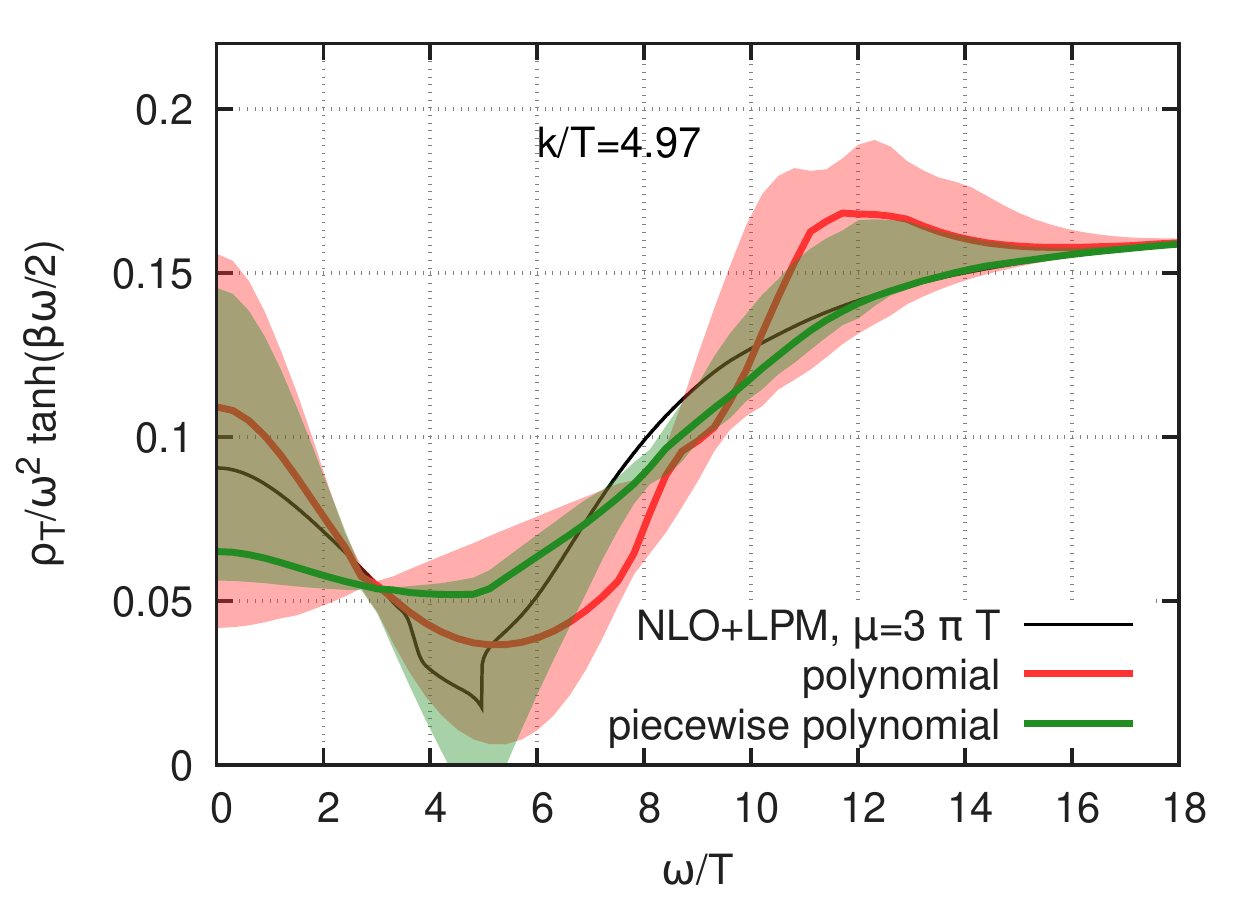}
	\caption{Reconstructed spectral functions using the NLO+LPM weak-coupling
        $G_\mathrm{T}/G_{00}$ mock correlator as an input.
	\label{fig:mock,nlo3,s=1}}
\end{figure}

\begin{figure}[t!]
	\includegraphics[scale=0.60]{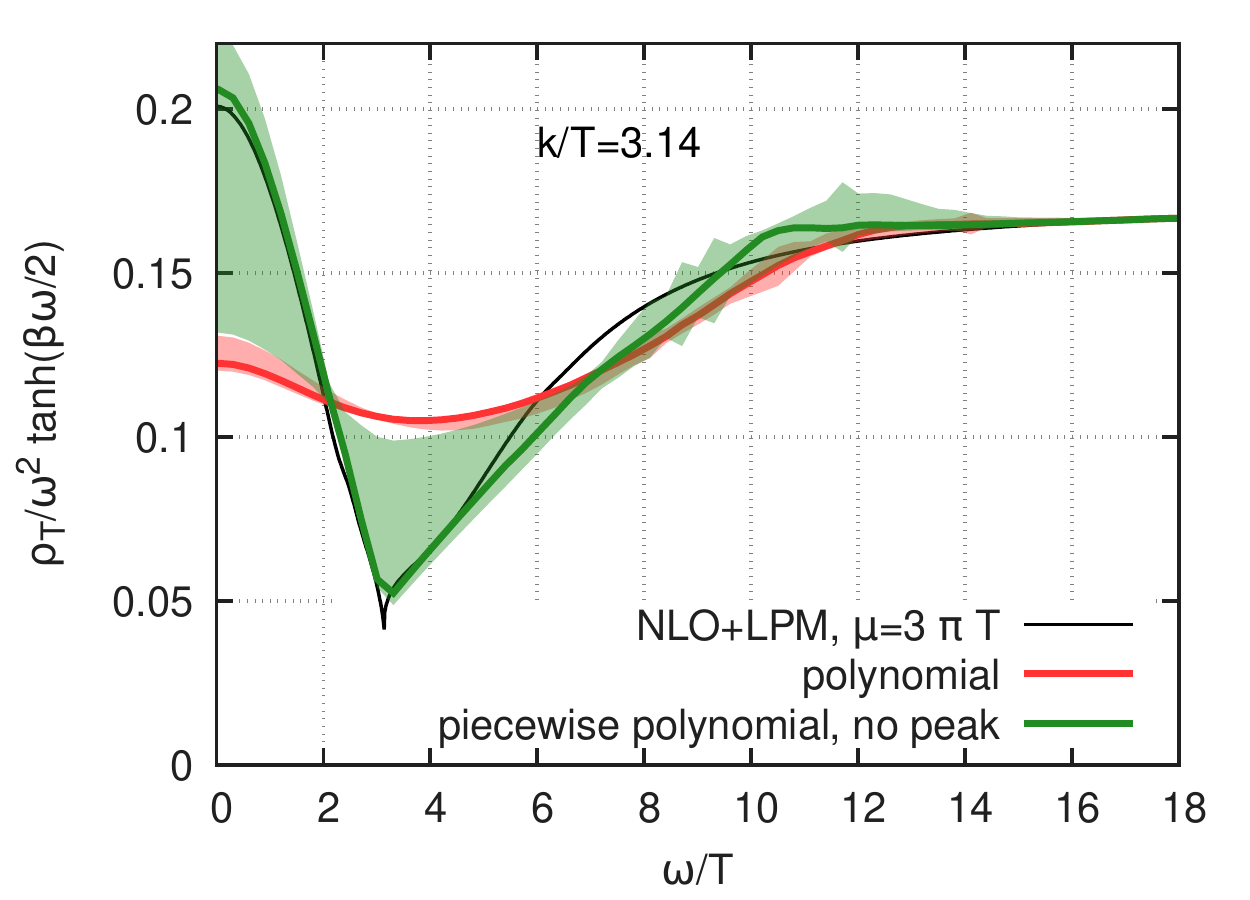}
	\includegraphics[scale=0.60]{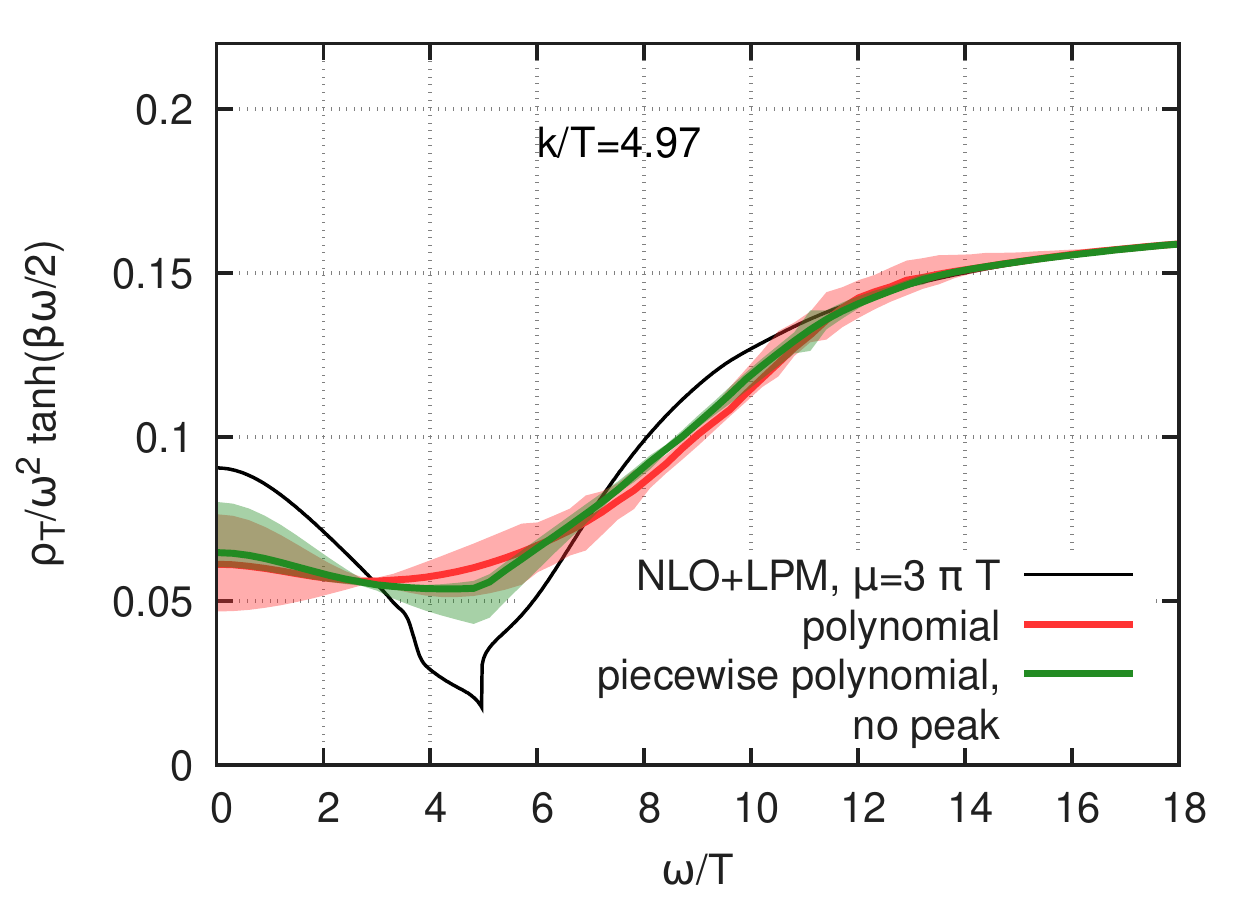}
	\caption{Reconstructed spectral functions using the NLO+LPM weak-coupling
        $G_\mathrm{T}/T^3$ mock correlator as an input.
	\label{fig:mock,nlo3,s=1,ncd}}
\end{figure}

When using the NLO weak-coupling mock correlator as an input for
the reconstruction via fitting, we applied some variations in the
fit setup, similar to the case of the analysis of the continuum 
extrapolated correlator (Sec.~\ref{sec:fitmethod}).
These include using either two or three fit parameters in the fit ansätze, 
changing the number of the correlator data points utilized in the fit,
and changing the features ($\om_0/T, \Delta/T$) of the matching
to the UV behavior.
Since when producing the mock correlator, we generated multivariate
Gaussian variables randomly, to eliminate a possible effect coming from the
random input, we used six different mock correlators generated with
the procedure discussed above.
The systematic error have been estimated using the AIC-weighted
histogram of the various fit results, and we used the mock jackknife
samples to estimate the statistical error of the mock analyses.
The same fit ansatz functions --- Eq.~(\ref{eq:fit,poly}) and
Eq.~(\ref{eq:fit,pipoly}) --- have been applied as in the main analysis.

We found that both fit ansätze perform reasonably well in 
the weak-coupling case when using the $G_\mathrm{T}/G_{00}$ data, see Fig.~\ref{fig:mock,nlo3,s=1}.
Namely, above $k/T \approx \pi$, both ans\"atze reproduce
the input spectral functions within errors in a wide range of 
$\omega/T$ values also at small and large frequencies.
The errors, however, are typically larger in this case for 
the polynomial ansatz ($\sim$ 10 -- 50\%, but even could be 100\%
at certain frequencies), than when using the $G_\mathrm{T}/T^3$ data
(errors < 2 -- 8\%).
At $\om = k$, however, the polynomial ansatz gives a much larger
value of $\rho$, i.e.\ a larger photon rate, using either $G_\mathrm{T}/G_{00}$
or $G_\mathrm{T}/T^3$.
The piecewise polynomial ansatz could reproduce the features of
the NLO+LPM mock data better, because it is capable of producing
a sharper dip at the light-cone.
With the help this ansatz, we get smaller photon rates at 
small momenta (below $k/T \approx \pi$), and larger photon
rates at larger momenta, but in this latter case with errors
spreading towards smaller values, typically covering the true
spectral function at the light-cone.
Using the $G_\mathrm{T}/T^3$ mock data, this ansatz --- similarly to
the polynomial ansatz --- also returns a larger photon rate
at all momenta, see e.g. Fig.~\ref{fig:mock,nlo3,s=1,ncd}.
In this case, the error covers the bottom of the dip only
at and below $k/T = \pi$.

The reproduction of the mock spectral function can be improved by
a certain amount by reducing the errors on the input mock correlators.
As Fig.~\ref{fig:mock,nlo,s=10} shows, accomodating an error reduction
factor of $s=10$, the piecewise polynomial ansatz could mimic the dip
at $\om = k$ at a more satisfactory level.
For comparison, see the right panel of Fig.~\ref{fig:mock,nlo3,s=1}
and also that of Fig.~\ref{fig:mock,nlo3,s=1,ncd}.
This observation shows that this particular ansatz having only a few
parameters has satisfactory expressiveness of reproducing a model spectral
function relevant to the physics discussed in this paper.
Conversely, it also indicates that the fact that we obtained less faithful spectral
function outcomes when not reducing the error is mainly due to the covariance
matrix and not due to a wrong choice of ansätze.

\begin{figure}[t!]
	\includegraphics[scale=0.60]{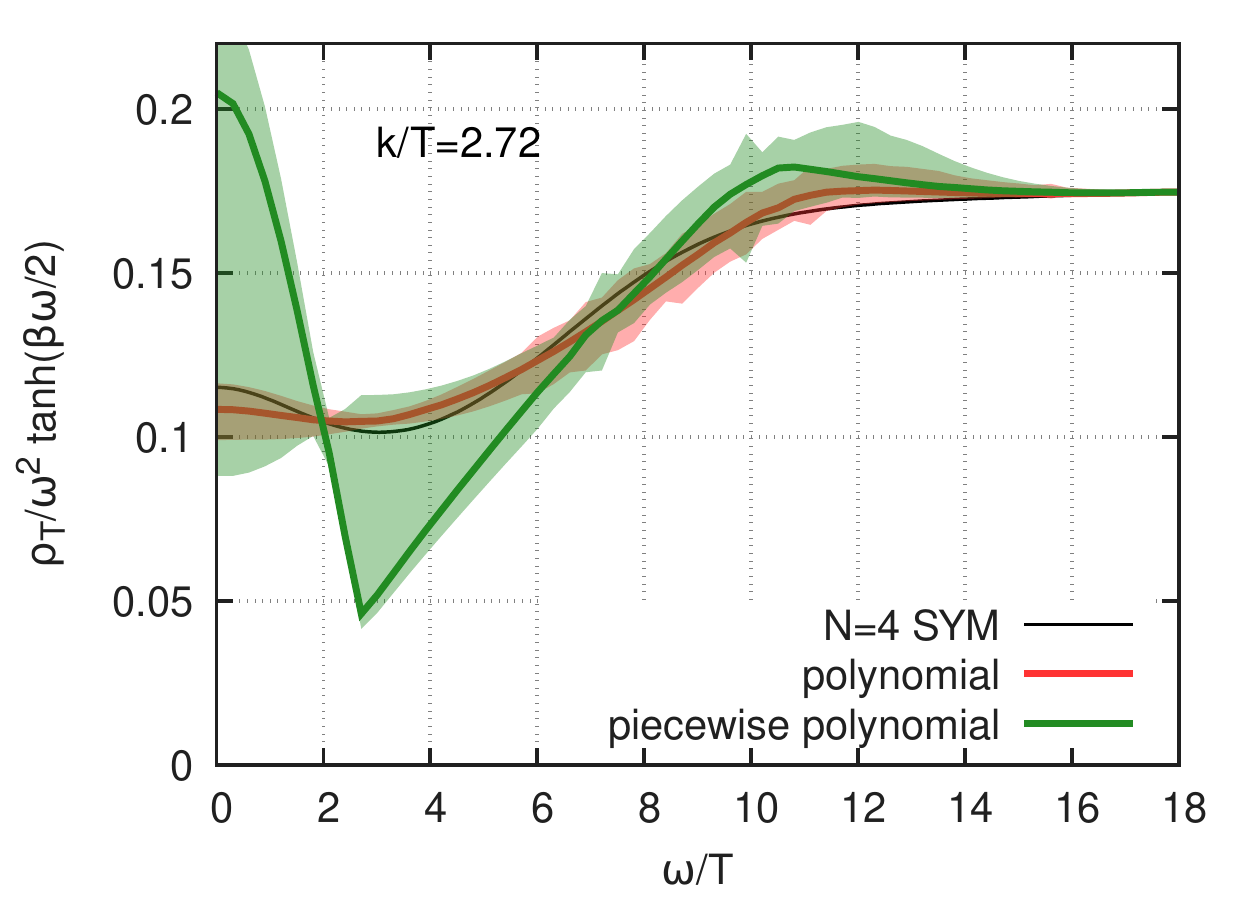}
	\includegraphics[scale=0.60]{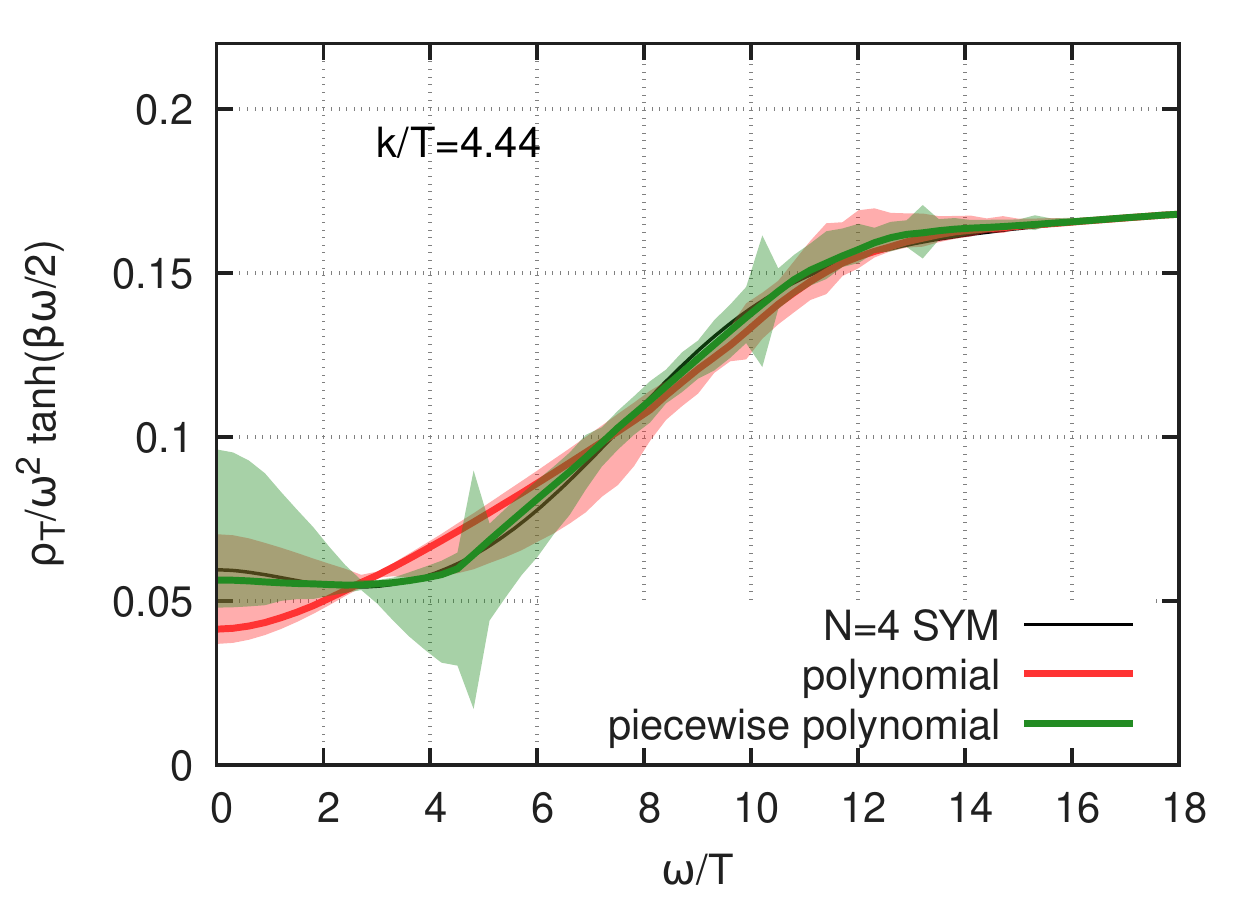}
	\caption{Reconstructed spectral functions using the strongly coupled $\mathcal{N}=4$ SYM
        theory mock correlator, $G_\mathrm{T}/G_{00}$, as an input.
	\label{fig:mock,adscft,s=1}}
\end{figure}

\begin{figure}[t!]
	\includegraphics[scale=0.60]{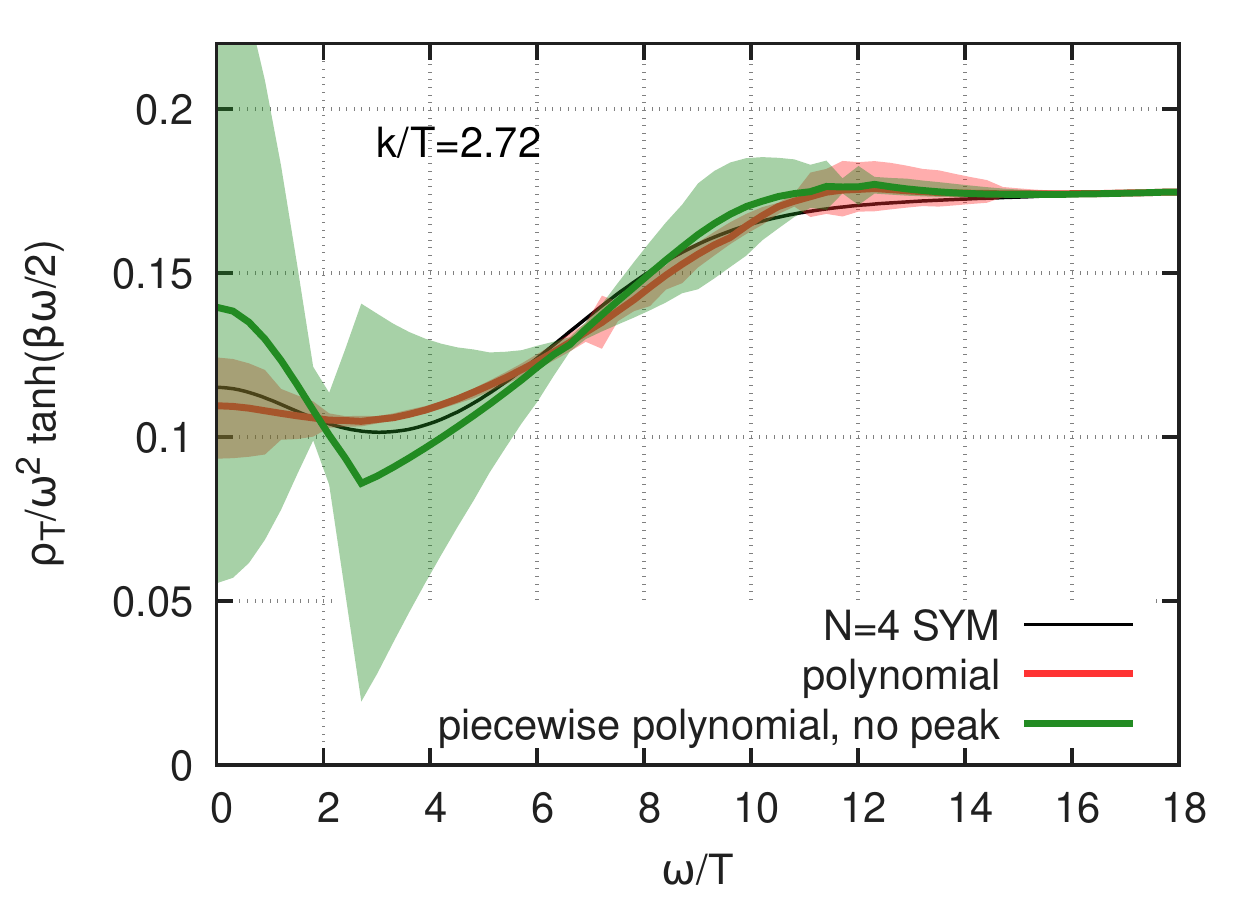}
	\includegraphics[scale=0.60]{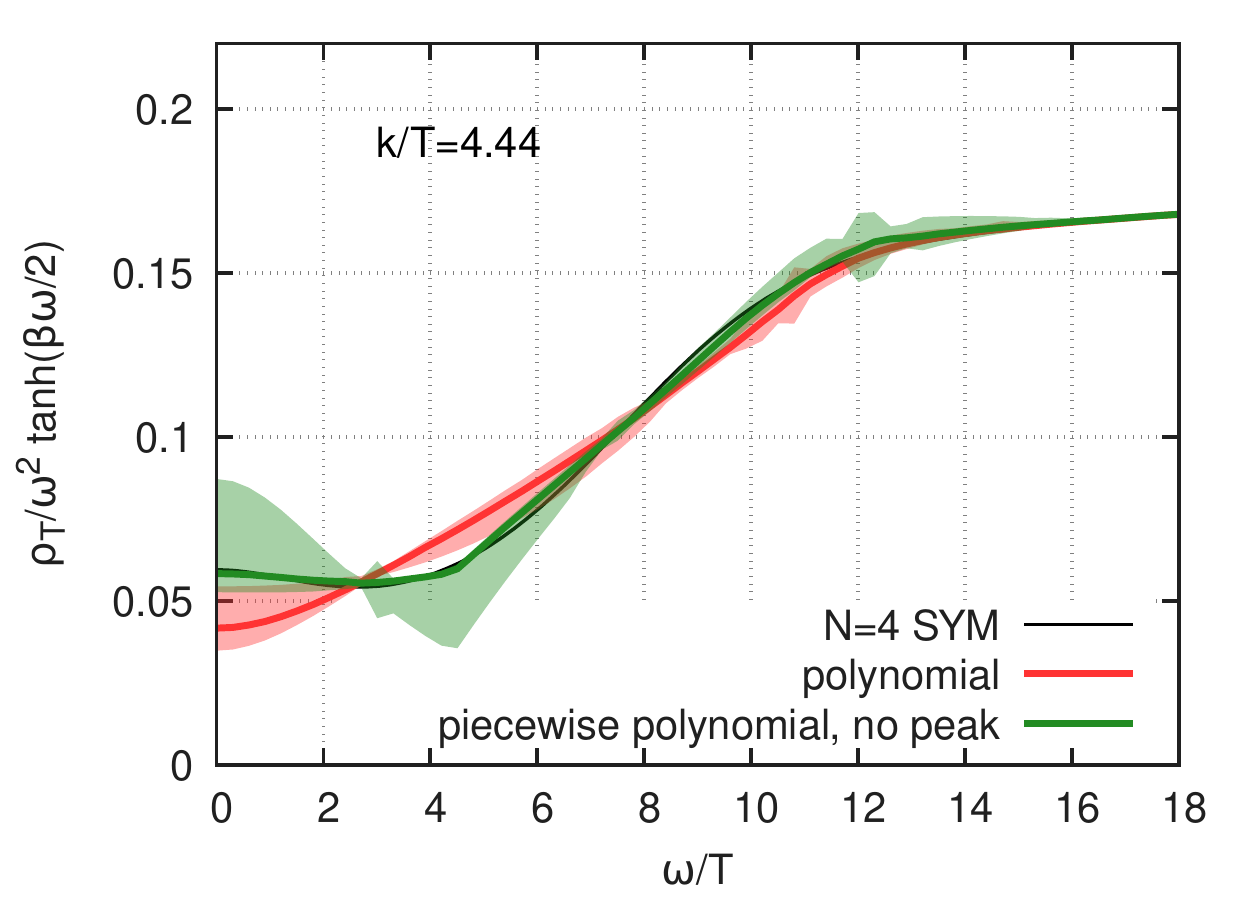}
	\caption{Reconstructed spectral functions using the strongly coupled $\mathcal{N}=4$ SYM
        theory mock correlator, $G_\mathrm{T}/T^3$, as an input.
	\label{fig:mock,adscft,s=1,ncd}}
\end{figure}

\begin{figure}[t!]
	\includegraphics[scale=0.60]{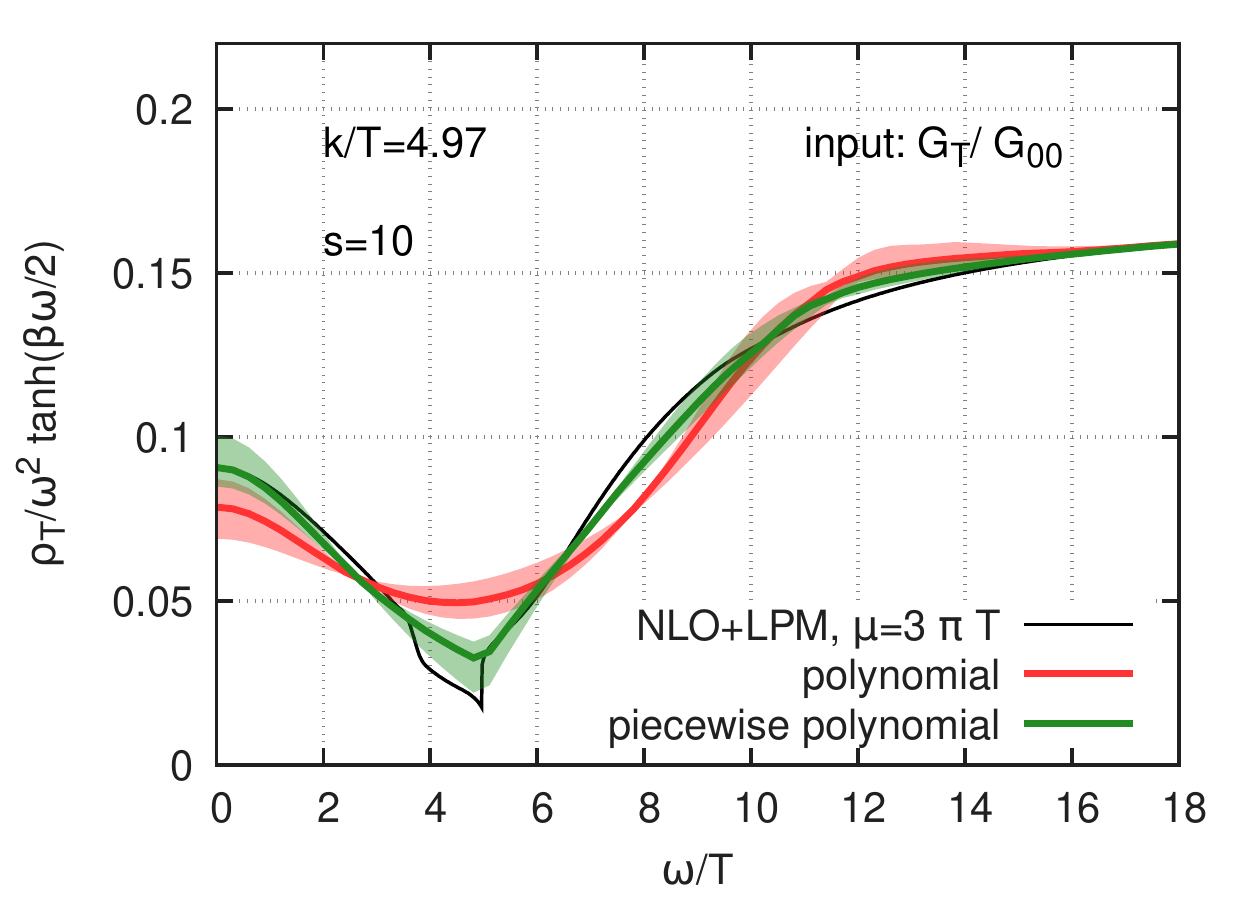}
	\includegraphics[scale=0.60]{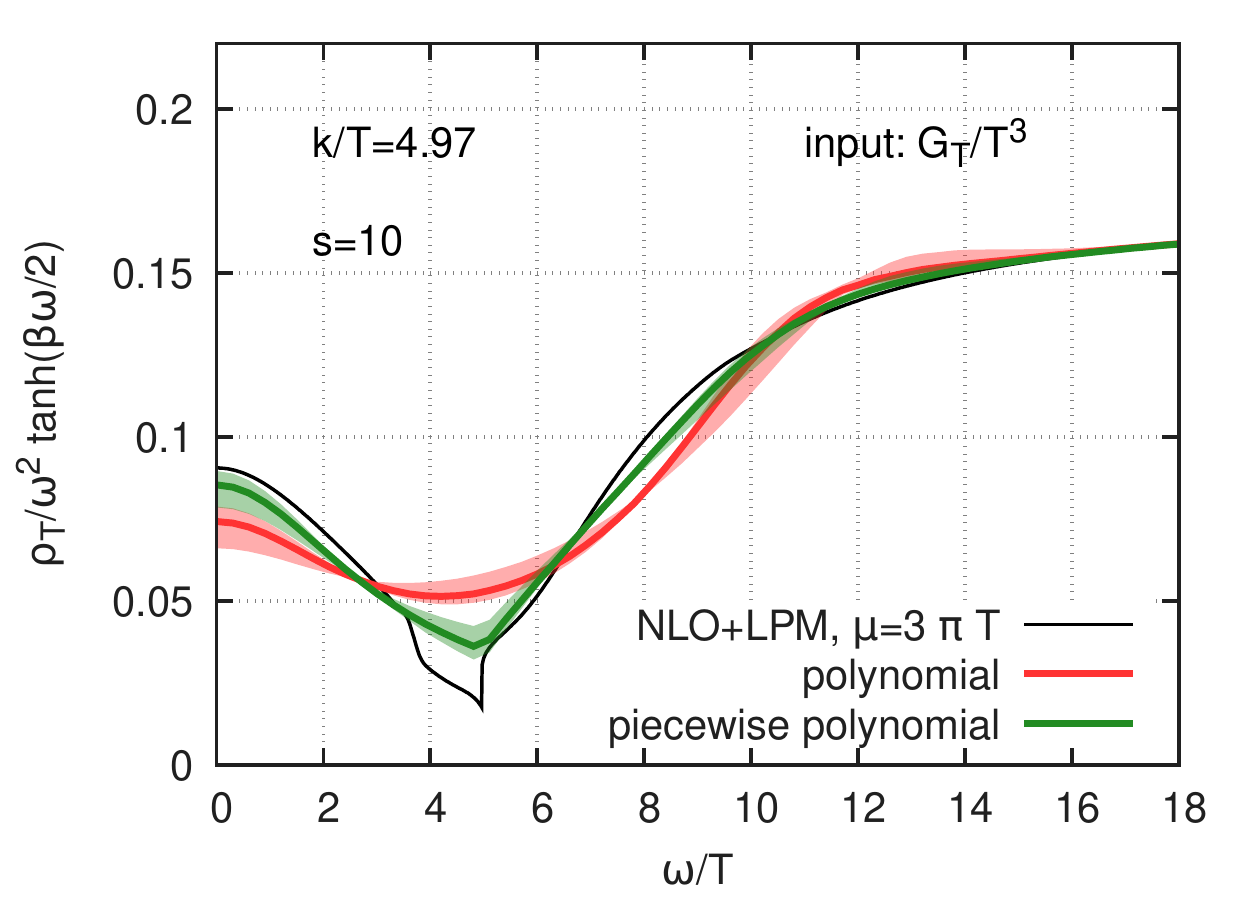}
	\caption{Reconstructed spectral functions using the NLO+LPM correlator 
        $G_\mathrm{T}/G_{00}$ ({\bf left panel}), or $G_\mathrm{T}/T^3$ ({\bf right panel}) as an input.
	The covariance matrix elements have been decreased by a factor of $s^2 = 100$.
	Having so small errors, the piecewise polynomial ansatz could better approximate
	the dip.
	\label{fig:mock,nlo,s=10}
	}
\end{figure}

Performing mock tests using the strongly coupled $\mathcal{N}=4$
SYM theory transverse correlators, the polynomial
ansatz adequately reproduces the spectral function even for small 
momenta with errors less than around 5 -- 10\% for frequencies
around and above $\om = k$, see Fig.~\ref{fig:mock,adscft,s=1} and
Fig.~\ref{fig:mock,adscft,s=1,ncd}.
As one goes for higher momenta, the reproduction gets a bit 
worse, but usually with errors included, one reaches the true
values of the input spectral function.
The photon rates given by this ansatz are always higher than the
true photon rate, either using the mock $G_\mathrm{T}/G_{00}$ or the $G_\mathrm{T}/T^3$
data, see Fig.~\ref{fig:mock,adscft,s=1} and Fig.~\ref{fig:mock,adscft,s=1,ncd},
respectively.
The deviation increases from about 2\% corresponding to smaller
momenta up to 25\% for high momenta.

For this model, the piecewise polynomial ansatz could also give
reasonably good results for high momenta, where it almost coincides
with the polynomial ansatz results.
For momenta, $k/T < 4.44$, it always gives a smaller photon rate,
not depending on using $G_\mathrm{T}/G_{00}$ or $G_\mathrm{T}/T^3$, but in the latter
case the central value is much closer to the true value.

The mock data also served as a numerical crosscheck for the Backus-Gilbert method,
both using $G_\mathrm{T}/G_{00}$ as well as $G_\mathrm{T}/T^3$.
As Fig.~\ref{fig:BG,mock,sYM} shows, the Backus-Gilbert method is capable
of reproducing the smeared, rescaled spectral function up to around $\om/T \sim 16$.

\begin{figure}[t!]
	\includegraphics[scale=0.60]{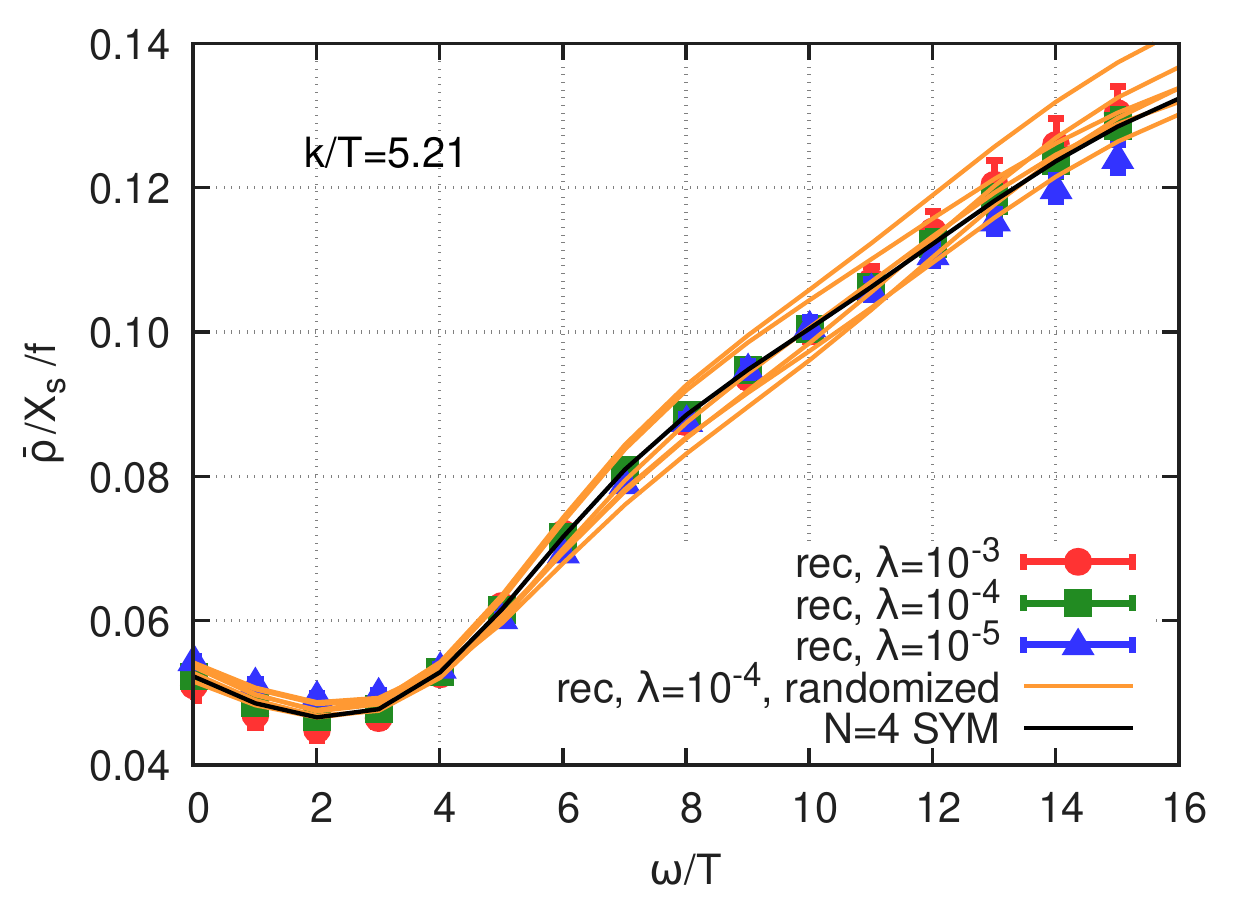}
	\caption{Reconstructed, filtered spectral functions using the $\mathcal{N}=4$
        SYM mock correlator,
		$\bar{G}^{(\mathrm{mock})}=G^{(\mathrm{mock})}_\mathrm{T}/G^{(\mathrm{mock})}_{00}$, as an input.
	The results obtained with the help of the 6 randomized mock correlators are
	shown in orange without including the statistical errors.
	These errors are comparable to the errors of the reconstructed $\hat{\rho}$ using
	the exact model data.
	\label{fig:BG,mock,sYM}}
\end{figure}


\newpage
\bibliography{pertGTfit}

\begin{thebibliography}{59}
\expandafter\ifx\csname natexlab\endcsname\relax\def\natexlab#1{#1}\fi
\expandafter\ifx\csname bibnamefont\endcsname\relax
  \def\bibnamefont#1{#1}\fi
\expandafter\ifx\csname bibfnamefont\endcsname\relax
  \def\bibfnamefont#1{#1}\fi
\expandafter\ifx\csname citenamefont\endcsname\relax
  \def\citenamefont#1{#1}\fi
\expandafter\ifx\csname url\endcsname\relax
  \def\url#1{\texttt{#1}}\fi
\expandafter\ifx\csname urlprefix\endcsname\relax\def\urlprefix{URL }\fi
\providecommand{\bibinfo}[2]{#2}
\providecommand{\eprint}[2][]{\url{#2}}

\bibitem[{\citenamefont{Feinberg}(1976)}]{Feinberg:1976ua}
\bibinfo{author}{\bibfnamefont{E.~L.} \bibnamefont{Feinberg}},
  \bibinfo{journal}{Nuovo Cim. A} \textbf{\bibinfo{volume}{34}},
  \bibinfo{pages}{391} (\bibinfo{year}{1976}).

\bibitem[{\citenamefont{Shuryak}(1978)}]{Shuryak:1978ij}
\bibinfo{author}{\bibfnamefont{E.~V.} \bibnamefont{Shuryak}},
  \bibinfo{journal}{Phys. Lett. B} \textbf{\bibinfo{volume}{78}},
  \bibinfo{pages}{150} (\bibinfo{year}{1978}).

\bibitem[{\citenamefont{Ferbel and Molzon}(1984)}]{Ferbel:1984ef}
\bibinfo{author}{\bibfnamefont{T.}~\bibnamefont{Ferbel}} \bibnamefont{and}
  \bibinfo{author}{\bibfnamefont{W.~R.} \bibnamefont{Molzon}},
  \bibinfo{journal}{Rev. Mod. Phys.} \textbf{\bibinfo{volume}{56}},
  \bibinfo{pages}{181} (\bibinfo{year}{1984}).

\bibitem[{\citenamefont{Cassing and Bratkovskaya}(1999)}]{Cassing:1999es}
\bibinfo{author}{\bibfnamefont{W.}~\bibnamefont{Cassing}} \bibnamefont{and}
  \bibinfo{author}{\bibfnamefont{E.~L.} \bibnamefont{Bratkovskaya}},
  \bibinfo{journal}{Phys. Rept.} \textbf{\bibinfo{volume}{308}},
  \bibinfo{pages}{65} (\bibinfo{year}{1999}).

\bibitem[{\citenamefont{David}(2020)}]{David:2019wpt}
\bibinfo{author}{\bibfnamefont{G.}~\bibnamefont{David}},
  \bibinfo{journal}{Rept. Prog. Phys.} \textbf{\bibinfo{volume}{83}},
  \bibinfo{pages}{046301} (\bibinfo{year}{2020}), \eprint{1907.08893}.

\bibitem[{\citenamefont{Adare et~al.}(2015)}]{Adare:2014fwh}
\bibinfo{author}{\bibfnamefont{A.}~\bibnamefont{Adare}} \bibnamefont{et~al.}
  (\bibinfo{collaboration}{PHENIX}), \bibinfo{journal}{Phys. Rev. C}
  \textbf{\bibinfo{volume}{91}}, \bibinfo{pages}{064904}
  (\bibinfo{year}{2015}), \eprint{1405.3940}.

\bibitem[{\citenamefont{Adamczyk et~al.}(2017)}]{STAR:2016use}
\bibinfo{author}{\bibfnamefont{L.}~\bibnamefont{Adamczyk}} \bibnamefont{et~al.}
  (\bibinfo{collaboration}{STAR}), \bibinfo{journal}{Phys. Lett. B}
  \textbf{\bibinfo{volume}{770}}, \bibinfo{pages}{451} (\bibinfo{year}{2017}),
  \eprint{1607.01447}.

\bibitem[{\citenamefont{Adam et~al.}(2016)}]{Adam:2015lda}
\bibinfo{author}{\bibfnamefont{J.}~\bibnamefont{Adam}} \bibnamefont{et~al.}
  (\bibinfo{collaboration}{ALICE}), \bibinfo{journal}{Phys. Lett. B}
  \textbf{\bibinfo{volume}{754}}, \bibinfo{pages}{235} (\bibinfo{year}{2016}),
  \eprint{1509.07324}.

\bibitem[{\citenamefont{Paquet et~al.}(2016)\citenamefont{Paquet, Shen,
  Denicol, Luzum, Schenke, Jeon, and Gale}}]{Paquet:2015lta}
\bibinfo{author}{\bibfnamefont{J.-F.} \bibnamefont{Paquet}},
  \bibinfo{author}{\bibfnamefont{C.}~\bibnamefont{Shen}},
  \bibinfo{author}{\bibfnamefont{G.~S.} \bibnamefont{Denicol}},
  \bibinfo{author}{\bibfnamefont{M.}~\bibnamefont{Luzum}},
  \bibinfo{author}{\bibfnamefont{B.}~\bibnamefont{Schenke}},
  \bibinfo{author}{\bibfnamefont{S.}~\bibnamefont{Jeon}}, \bibnamefont{and}
  \bibinfo{author}{\bibfnamefont{C.}~\bibnamefont{Gale}},
  \bibinfo{journal}{Phys. Rev. C} \textbf{\bibinfo{volume}{93}},
  \bibinfo{pages}{044906} (\bibinfo{year}{2016}), \eprint{1509.06738}.

\bibitem[{\citenamefont{Gale et~al.}(2021)\citenamefont{Gale, Paquet, Schenke,
  and Shen}}]{Gale:2021zlc}
\bibinfo{author}{\bibfnamefont{C.}~\bibnamefont{Gale}},
  \bibinfo{author}{\bibfnamefont{J.-F.} \bibnamefont{Paquet}},
  \bibinfo{author}{\bibfnamefont{B.}~\bibnamefont{Schenke}}, \bibnamefont{and}
  \bibinfo{author}{\bibfnamefont{C.}~\bibnamefont{Shen}}
  (\bibinfo{year}{2021}), \eprint{2106.11216}.

\bibitem[{\citenamefont{Arnold et~al.}(2001)\citenamefont{Arnold, Moore, and
  Yaffe}}]{Arnold:2001ms}
\bibinfo{author}{\bibfnamefont{P.~B.} \bibnamefont{Arnold}},
  \bibinfo{author}{\bibfnamefont{G.~D.} \bibnamefont{Moore}}, \bibnamefont{and}
  \bibinfo{author}{\bibfnamefont{L.~G.} \bibnamefont{Yaffe}},
  \bibinfo{journal}{JHEP} \textbf{\bibinfo{volume}{12}}, \bibinfo{pages}{009}
  (\bibinfo{year}{2001}), \eprint{hep-ph/0111107}.

\bibitem[{\citenamefont{Ghiglieri et~al.}(2013)\citenamefont{Ghiglieri, Hong,
  Kurkela, Lu, Moore et~al.}}]{Ghiglieri:2013gia}
\bibinfo{author}{\bibfnamefont{J.}~\bibnamefont{Ghiglieri}},
  \bibinfo{author}{\bibfnamefont{J.}~\bibnamefont{Hong}},
  \bibinfo{author}{\bibfnamefont{A.}~\bibnamefont{Kurkela}},
  \bibinfo{author}{\bibfnamefont{E.}~\bibnamefont{Lu}},
  \bibinfo{author}{\bibfnamefont{G.~D.} \bibnamefont{Moore}},
  \bibnamefont{et~al.}, \bibinfo{journal}{JHEP}
  \textbf{\bibinfo{volume}{1305}}, \bibinfo{pages}{010} (\bibinfo{year}{2013}),
  \eprint{1302.5970}.

\bibitem[{\citenamefont{Aarts and Martinez~Resco}(2002)}]{Aarts:2002cc}
\bibinfo{author}{\bibfnamefont{G.}~\bibnamefont{Aarts}} \bibnamefont{and}
  \bibinfo{author}{\bibfnamefont{J.~M.} \bibnamefont{Martinez~Resco}},
  \bibinfo{journal}{JHEP} \textbf{\bibinfo{volume}{04}}, \bibinfo{pages}{053}
  (\bibinfo{year}{2002}), \eprint{hep-ph/0203177}.

\bibitem[{\citenamefont{Teaney}(2006)}]{Teaney:2006nc}
\bibinfo{author}{\bibfnamefont{D.}~\bibnamefont{Teaney}},
  \bibinfo{journal}{Phys. Rev. D} \textbf{\bibinfo{volume}{74}},
  \bibinfo{pages}{045025} (\bibinfo{year}{2006}), \eprint{hep-ph/0602044}.

\bibitem[{\citenamefont{Meyer}(2011)}]{Meyer:2011gj}
\bibinfo{author}{\bibfnamefont{H.~B.} \bibnamefont{Meyer}},
  \bibinfo{journal}{Eur. Phys. J. A} \textbf{\bibinfo{volume}{47}},
  \bibinfo{pages}{86} (\bibinfo{year}{2011}), \eprint{1104.3708}.

\bibitem[{\citenamefont{Bors\'anyi et~al.}(2018)\citenamefont{Bors\'anyi,
  Fodor, Giordano, Katz, Pasztor, Ratti, Sch\"afer, Szabo, and
  T\'oth}}]{Borsanyi:2018srz}
\bibinfo{author}{\bibfnamefont{S.}~\bibnamefont{Bors\'anyi}},
  \bibinfo{author}{\bibfnamefont{Z.}~\bibnamefont{Fodor}},
  \bibinfo{author}{\bibfnamefont{M.}~\bibnamefont{Giordano}},
  \bibinfo{author}{\bibfnamefont{S.~D.} \bibnamefont{Katz}},
  \bibinfo{author}{\bibfnamefont{A.}~\bibnamefont{Pasztor}},
  \bibinfo{author}{\bibfnamefont{C.}~\bibnamefont{Ratti}},
  \bibinfo{author}{\bibfnamefont{A.}~\bibnamefont{Sch\"afer}},
  \bibinfo{author}{\bibfnamefont{K.~K.} \bibnamefont{Szabo}}, \bibnamefont{and}
  \bibinfo{author}{\bibfnamefont{B.~C.} \bibnamefont{T\'oth}},
  \bibinfo{journal}{Phys. Rev. D} \textbf{\bibinfo{volume}{98}},
  \bibinfo{pages}{014512} (\bibinfo{year}{2018}), \eprint{1802.07718}.

\bibitem[{\citenamefont{Ding et~al.}(2011)\citenamefont{Ding, Francis,
  Kaczmarek, Karsch, Laermann et~al.}}]{Ding:2010ga}
\bibinfo{author}{\bibfnamefont{H.-T.} \bibnamefont{Ding}},
  \bibinfo{author}{\bibfnamefont{A.}~\bibnamefont{Francis}},
  \bibinfo{author}{\bibfnamefont{O.}~\bibnamefont{Kaczmarek}},
  \bibinfo{author}{\bibfnamefont{F.}~\bibnamefont{Karsch}},
  \bibinfo{author}{\bibfnamefont{E.}~\bibnamefont{Laermann}},
  \bibnamefont{et~al.}, \bibinfo{journal}{Phys. Rev. D}
  \textbf{\bibinfo{volume}{83}}, \bibinfo{pages}{034504}
  (\bibinfo{year}{2011}), \eprint{1012.4963}.

\bibitem[{\citenamefont{Brandt et~al.}(2013)\citenamefont{Brandt, Francis,
  Meyer, and Wittig}}]{Brandt:2012jc}
\bibinfo{author}{\bibfnamefont{B.~B.} \bibnamefont{Brandt}},
  \bibinfo{author}{\bibfnamefont{A.}~\bibnamefont{Francis}},
  \bibinfo{author}{\bibfnamefont{H.~B.} \bibnamefont{Meyer}}, \bibnamefont{and}
  \bibinfo{author}{\bibfnamefont{H.}~\bibnamefont{Wittig}},
  \bibinfo{journal}{JHEP} \textbf{\bibinfo{volume}{03}}, \bibinfo{pages}{100}
  (\bibinfo{year}{2013}), \eprint{1212.4200}.

\bibitem[{\citenamefont{Amato et~al.}(2013)\citenamefont{Amato, Aarts, Allton,
  Giudice, Hands et~al.}}]{Amato:2013naa}
\bibinfo{author}{\bibfnamefont{A.}~\bibnamefont{Amato}},
  \bibinfo{author}{\bibfnamefont{G.}~\bibnamefont{Aarts}},
  \bibinfo{author}{\bibfnamefont{C.}~\bibnamefont{Allton}},
  \bibinfo{author}{\bibfnamefont{P.}~\bibnamefont{Giudice}},
  \bibinfo{author}{\bibfnamefont{S.}~\bibnamefont{Hands}},
  \bibnamefont{et~al.}, \bibinfo{journal}{Phys. Rev. Lett.}
  \textbf{\bibinfo{volume}{111}}, \bibinfo{pages}{172001}
  (\bibinfo{year}{2013}), \eprint{1307.6763}.

\bibitem[{\citenamefont{Aarts et~al.}(2015)\citenamefont{Aarts, Allton, Amato,
  Giudice, Hands, and Skullerud}}]{Aarts:2014nba}
\bibinfo{author}{\bibfnamefont{G.}~\bibnamefont{Aarts}},
  \bibinfo{author}{\bibfnamefont{C.}~\bibnamefont{Allton}},
  \bibinfo{author}{\bibfnamefont{A.}~\bibnamefont{Amato}},
  \bibinfo{author}{\bibfnamefont{P.}~\bibnamefont{Giudice}},
  \bibinfo{author}{\bibfnamefont{S.}~\bibnamefont{Hands}}, \bibnamefont{and}
  \bibinfo{author}{\bibfnamefont{J.-I.} \bibnamefont{Skullerud}},
  \bibinfo{journal}{JHEP} \textbf{\bibinfo{volume}{02}}, \bibinfo{pages}{186}
  (\bibinfo{year}{2015}), \eprint{1412.6411}.

\bibitem[{\citenamefont{Brandt et~al.}(2016{\natexlab{a}})\citenamefont{Brandt,
  Francis, Jaeger, and Meyer}}]{Brandt:2015aqk}
\bibinfo{author}{\bibfnamefont{B.~B.} \bibnamefont{Brandt}},
  \bibinfo{author}{\bibfnamefont{A.}~\bibnamefont{Francis}},
  \bibinfo{author}{\bibfnamefont{B.}~\bibnamefont{Jaeger}}, \bibnamefont{and}
  \bibinfo{author}{\bibfnamefont{H.~B.} \bibnamefont{Meyer}},
  \bibinfo{journal}{Phys. Rev. D} \textbf{\bibinfo{volume}{93}},
  \bibinfo{pages}{054510} (\bibinfo{year}{2016}{\natexlab{a}}),
  \eprint{1512.07249}.

\bibitem[{\citenamefont{Ding et~al.}(2016)\citenamefont{Ding, Kaczmarek, and
  Meyer}}]{Ding:2016hua}
\bibinfo{author}{\bibfnamefont{H.-T.} \bibnamefont{Ding}},
  \bibinfo{author}{\bibfnamefont{O.}~\bibnamefont{Kaczmarek}},
  \bibnamefont{and} \bibinfo{author}{\bibfnamefont{F.}~\bibnamefont{Meyer}},
  \bibinfo{journal}{Phys. Rev. D} \textbf{\bibinfo{volume}{94}},
  \bibinfo{pages}{034504} (\bibinfo{year}{2016}), \eprint{1604.06712}.

\bibitem[{\citenamefont{Ghiglieri et~al.}(2016)\citenamefont{Ghiglieri,
  Kaczmarek, Laine, and Meyer}}]{Ghiglieri:2016tvj}
\bibinfo{author}{\bibfnamefont{J.}~\bibnamefont{Ghiglieri}},
  \bibinfo{author}{\bibfnamefont{O.}~\bibnamefont{Kaczmarek}},
  \bibinfo{author}{\bibfnamefont{M.}~\bibnamefont{Laine}}, \bibnamefont{and}
  \bibinfo{author}{\bibfnamefont{F.}~\bibnamefont{Meyer}},
  \bibinfo{journal}{Phys. Rev. D} \textbf{\bibinfo{volume}{94}},
  \bibinfo{pages}{016005} (\bibinfo{year}{2016}), \eprint{1604.07544}.

\bibitem[{\citenamefont{C\`e et~al.}(2020)\citenamefont{C\`e, Harris, Meyer,
  Steinberg, and Toniato}}]{Ce:2020tmx}
\bibinfo{author}{\bibfnamefont{M.}~\bibnamefont{C\`e}},
  \bibinfo{author}{\bibfnamefont{T.}~\bibnamefont{Harris}},
  \bibinfo{author}{\bibfnamefont{H.~B.} \bibnamefont{Meyer}},
  \bibinfo{author}{\bibfnamefont{A.}~\bibnamefont{Steinberg}},
  \bibnamefont{and} \bibinfo{author}{\bibfnamefont{A.}~\bibnamefont{Toniato}},
  \bibinfo{journal}{Phys. Rev. D} \textbf{\bibinfo{volume}{102}},
  \bibinfo{pages}{091501} (\bibinfo{year}{2020}), \eprint{2001.03368}.

\bibitem[{\citenamefont{Backus and Gilbert}(1968)}]{Backus:1968abc}
\bibinfo{author}{\bibfnamefont{G.}~\bibnamefont{Backus}} \bibnamefont{and}
  \bibinfo{author}{\bibfnamefont{F.}~\bibnamefont{Gilbert}},
  \bibinfo{journal}{Geophys. J. Int.} \textbf{\bibinfo{volume}{16}},
  \bibinfo{pages}{169} (\bibinfo{year}{1968}).

\bibitem[{\citenamefont{Backus and Gilbert}(1970)}]{Backus:1970}
\bibinfo{author}{\bibfnamefont{G.}~\bibnamefont{Backus}} \bibnamefont{and}
  \bibinfo{author}{\bibfnamefont{F.}~\bibnamefont{Gilbert}},
  \bibinfo{journal}{Philos. T. R. Soc. A} \textbf{\bibinfo{volume}{266}},
  \bibinfo{pages}{123} (\bibinfo{year}{1970}).

\bibitem[{\citenamefont{McLerran and Toimela}(1985)}]{McLerran:1985abc}
\bibinfo{author}{\bibfnamefont{L.~D.} \bibnamefont{McLerran}} \bibnamefont{and}
  \bibinfo{author}{\bibfnamefont{T.}~\bibnamefont{Toimela}},
  \bibinfo{journal}{Phys. Rev. D} \textbf{\bibinfo{volume}{31}},
  \bibinfo{pages}{545} (\bibinfo{year}{1985}).

\bibitem[{\citenamefont{Brandt et~al.}(2018)\citenamefont{Brandt, Francis,
  Harris, Meyer, and Steinberg}}]{Brandt:2017vgl}
\bibinfo{author}{\bibfnamefont{B.~B.} \bibnamefont{Brandt}},
  \bibinfo{author}{\bibfnamefont{A.}~\bibnamefont{Francis}},
  \bibinfo{author}{\bibfnamefont{T.}~\bibnamefont{Harris}},
  \bibinfo{author}{\bibfnamefont{H.~B.} \bibnamefont{Meyer}}, \bibnamefont{and}
  \bibinfo{author}{\bibfnamefont{A.}~\bibnamefont{Steinberg}},
  \bibinfo{journal}{EPJ Web Conf.} \textbf{\bibinfo{volume}{175}},
  \bibinfo{pages}{07044} (\bibinfo{year}{2018}), \eprint{1710.07050}.

\bibitem[{\citenamefont{Brandt et~al.}(2019)\citenamefont{Brandt, C\`e,
  Francis, Harris, Meyer, Steinberg, and Toniato}}]{Brandt:2019shg}
\bibinfo{author}{\bibfnamefont{B.~B.} \bibnamefont{Brandt}},
  \bibinfo{author}{\bibfnamefont{M.}~\bibnamefont{C\`e}},
  \bibinfo{author}{\bibfnamefont{A.}~\bibnamefont{Francis}},
  \bibinfo{author}{\bibfnamefont{T.}~\bibnamefont{Harris}},
  \bibinfo{author}{\bibfnamefont{H.~B.} \bibnamefont{Meyer}},
  \bibinfo{author}{\bibfnamefont{A.}~\bibnamefont{Steinberg}},
  \bibnamefont{and} \bibinfo{author}{\bibfnamefont{A.}~\bibnamefont{Toniato}},
  \bibinfo{journal}{PoS} \textbf{\bibinfo{volume}{LATTICE2019}},
  \bibinfo{pages}{225} (\bibinfo{year}{2019}), \eprint{1912.00292}.

\bibitem[{\citenamefont{Hong and Teaney}(2010)}]{Hong:2010at}
\bibinfo{author}{\bibfnamefont{J.}~\bibnamefont{Hong}} \bibnamefont{and}
  \bibinfo{author}{\bibfnamefont{D.}~\bibnamefont{Teaney}},
  \bibinfo{journal}{Phys. Rev. C} \textbf{\bibinfo{volume}{82}},
  \bibinfo{pages}{044908} (\bibinfo{year}{2010}), \eprint{1003.0699}.

\bibitem[{\citenamefont{Baier et~al.}(1988)\citenamefont{Baier, Pire, and
  Schiff}}]{Baier:1988xv}
\bibinfo{author}{\bibfnamefont{R.}~\bibnamefont{Baier}},
  \bibinfo{author}{\bibfnamefont{B.}~\bibnamefont{Pire}}, \bibnamefont{and}
  \bibinfo{author}{\bibfnamefont{D.}~\bibnamefont{Schiff}},
  \bibinfo{journal}{Phys. Rev. D} \textbf{\bibinfo{volume}{38}},
  \bibinfo{pages}{2814} (\bibinfo{year}{1988}).

\bibitem[{\citenamefont{Aarts and Martinez~Resco}(2005)}]{Aarts:2005hg}
\bibinfo{author}{\bibfnamefont{G.}~\bibnamefont{Aarts}} \bibnamefont{and}
  \bibinfo{author}{\bibfnamefont{J.~M.} \bibnamefont{Martinez~Resco}},
  \bibinfo{journal}{Nucl. Phys. B} \textbf{\bibinfo{volume}{726}},
  \bibinfo{pages}{93} (\bibinfo{year}{2005}), \eprint{hep-lat/0507004}.

\bibitem[{\citenamefont{Laine}(2013)}]{Laine:2013vma}
\bibinfo{author}{\bibfnamefont{M.}~\bibnamefont{Laine}},
  \bibinfo{journal}{JHEP} \textbf{\bibinfo{volume}{11}}, \bibinfo{pages}{120}
  (\bibinfo{year}{2013}), \eprint{1310.0164}.

\bibitem[{\citenamefont{Ghisoiu and Laine}(2014)}]{Ghisoiu:2014mha}
\bibinfo{author}{\bibfnamefont{I.}~\bibnamefont{Ghisoiu}} \bibnamefont{and}
  \bibinfo{author}{\bibfnamefont{M.}~\bibnamefont{Laine}},
  \bibinfo{journal}{JHEP} \textbf{\bibinfo{volume}{10}}, \bibinfo{pages}{083}
  (\bibinfo{year}{2014}), \eprint{1407.7955}.

\bibitem[{\citenamefont{Jackson}(2019)}]{Jackson:2019mop}
\bibinfo{author}{\bibfnamefont{G.}~\bibnamefont{Jackson}},
  \bibinfo{journal}{Phys. Rev. D} \textbf{\bibinfo{volume}{100}},
  \bibinfo{pages}{116019} (\bibinfo{year}{2019}), \eprint{1910.07552}.

\bibitem[{\citenamefont{Jackson and Laine}(2019)}]{Jackson:2019yao}
\bibinfo{author}{\bibfnamefont{G.}~\bibnamefont{Jackson}} \bibnamefont{and}
  \bibinfo{author}{\bibfnamefont{M.}~\bibnamefont{Laine}},
  \bibinfo{journal}{JHEP} \textbf{\bibinfo{volume}{11}}, \bibinfo{pages}{144}
  (\bibinfo{year}{2019}), \eprint{1910.09567}.

\bibitem[{\citenamefont{Aurenche et~al.}(2000)\citenamefont{Aurenche, Gelis,
  and Zaraket}}]{Aurenche:2000gf}
\bibinfo{author}{\bibfnamefont{P.}~\bibnamefont{Aurenche}},
  \bibinfo{author}{\bibfnamefont{F.}~\bibnamefont{Gelis}}, \bibnamefont{and}
  \bibinfo{author}{\bibfnamefont{H.}~\bibnamefont{Zaraket}},
  \bibinfo{journal}{Phys. Rev. D} \textbf{\bibinfo{volume}{62}},
  \bibinfo{pages}{096012} (\bibinfo{year}{2000}), \eprint{hep-ph/0003326}.

\bibitem[{\citenamefont{Aurenche et~al.}(2002)\citenamefont{Aurenche, Gelis,
  Moore, and Zaraket}}]{Aurenche:2002wq}
\bibinfo{author}{\bibfnamefont{P.}~\bibnamefont{Aurenche}},
  \bibinfo{author}{\bibfnamefont{F.}~\bibnamefont{Gelis}},
  \bibinfo{author}{\bibfnamefont{G.~D.} \bibnamefont{Moore}}, \bibnamefont{and}
  \bibinfo{author}{\bibfnamefont{H.}~\bibnamefont{Zaraket}},
  \bibinfo{journal}{JHEP} \textbf{\bibinfo{volume}{12}}, \bibinfo{pages}{006}
  (\bibinfo{year}{2002}), \eprint{hep-ph/0211036}.

\bibitem[{\citenamefont{Carrington et~al.}(2008)\citenamefont{Carrington,
  Gynther, and Aurenche}}]{Carrington:2007gt}
\bibinfo{author}{\bibfnamefont{M.~E.} \bibnamefont{Carrington}},
  \bibinfo{author}{\bibfnamefont{A.}~\bibnamefont{Gynther}}, \bibnamefont{and}
  \bibinfo{author}{\bibfnamefont{P.}~\bibnamefont{Aurenche}},
  \bibinfo{journal}{Phys. Rev. D} \textbf{\bibinfo{volume}{77}},
  \bibinfo{pages}{045035} (\bibinfo{year}{2008}), \eprint{0711.3943}.

\bibitem[{\citenamefont{Caron-Huot et~al.}(2006)\citenamefont{Caron-Huot,
  Kovtun, Moore, Starinets, and Yaffe}}]{Caron-Huot:2006pee}
\bibinfo{author}{\bibfnamefont{S.}~\bibnamefont{Caron-Huot}},
  \bibinfo{author}{\bibfnamefont{P.}~\bibnamefont{Kovtun}},
  \bibinfo{author}{\bibfnamefont{G.~D.} \bibnamefont{Moore}},
  \bibinfo{author}{\bibfnamefont{A.}~\bibnamefont{Starinets}},
  \bibnamefont{and} \bibinfo{author}{\bibfnamefont{L.~G.} \bibnamefont{Yaffe}},
  \bibinfo{journal}{JHEP} \textbf{\bibinfo{volume}{12}}, \bibinfo{pages}{015}
  (\bibinfo{year}{2006}), \eprint{hep-th/0607237}.

\bibitem[{\citenamefont{Fritzsch et~al.}(2012)\citenamefont{Fritzsch, Knechtli,
  Leder, Marinkovic, Schaefer, Sommer, and Virotta}}]{Fritzsch:2012wq}
\bibinfo{author}{\bibfnamefont{P.}~\bibnamefont{Fritzsch}},
  \bibinfo{author}{\bibfnamefont{F.}~\bibnamefont{Knechtli}},
  \bibinfo{author}{\bibfnamefont{B.}~\bibnamefont{Leder}},
  \bibinfo{author}{\bibfnamefont{M.}~\bibnamefont{Marinkovic}},
  \bibinfo{author}{\bibfnamefont{S.}~\bibnamefont{Schaefer}},
  \bibinfo{author}{\bibfnamefont{R.}~\bibnamefont{Sommer}}, \bibnamefont{and}
  \bibinfo{author}{\bibfnamefont{F.}~\bibnamefont{Virotta}},
  \bibinfo{journal}{Nucl. Phys. B} \textbf{\bibinfo{volume}{865}},
  \bibinfo{pages}{397} (\bibinfo{year}{2012}), \eprint{1205.5380}.

\bibitem[{\citenamefont{Brandt et~al.}(2016{\natexlab{b}})\citenamefont{Brandt,
  Francis, Meyer, Philipsen, Robaina, and Wittig}}]{Brandt:2016daq}
\bibinfo{author}{\bibfnamefont{B.~B.} \bibnamefont{Brandt}},
  \bibinfo{author}{\bibfnamefont{A.}~\bibnamefont{Francis}},
  \bibinfo{author}{\bibfnamefont{H.~B.} \bibnamefont{Meyer}},
  \bibinfo{author}{\bibfnamefont{O.}~\bibnamefont{Philipsen}},
  \bibinfo{author}{\bibfnamefont{D.}~\bibnamefont{Robaina}}, \bibnamefont{and}
  \bibinfo{author}{\bibfnamefont{H.}~\bibnamefont{Wittig}},
  \bibinfo{journal}{JHEP} \textbf{\bibinfo{volume}{12}}, \bibinfo{pages}{158}
  (\bibinfo{year}{2016}{\natexlab{b}}), \eprint{1608.06882}.

\bibitem[{\citenamefont{Engel et~al.}(2015)\citenamefont{Engel, Giusti,
  Lottini, and Sommer}}]{Engel:2014cka}
\bibinfo{author}{\bibfnamefont{G.~P.} \bibnamefont{Engel}},
  \bibinfo{author}{\bibfnamefont{L.}~\bibnamefont{Giusti}},
  \bibinfo{author}{\bibfnamefont{S.}~\bibnamefont{Lottini}}, \bibnamefont{and}
  \bibinfo{author}{\bibfnamefont{R.}~\bibnamefont{Sommer}},
  \bibinfo{journal}{Phys. Rev. Lett.} \textbf{\bibinfo{volume}{114}},
  \bibinfo{pages}{112001} (\bibinfo{year}{2015}), \eprint{1406.4987}.

\bibitem[{\citenamefont{Steinberg}(2021)}]{Steinberg:2021bgr}
\bibinfo{author}{\bibfnamefont{A.}~\bibnamefont{Steinberg}}, Ph.D. thesis,
  \bibinfo{school}{Mainz U.} (\bibinfo{year}{2021}).

\bibitem[{\citenamefont{Akima}(1970)}]{Akima:1970abc}
\bibinfo{author}{\bibfnamefont{H.}~\bibnamefont{Akima}}, \bibinfo{journal}{J.
  ACM} \textbf{\bibinfo{volume}{17}}, \bibinfo{pages}{589–602}
  (\bibinfo{year}{1970}).

\bibitem[{\citenamefont{{Steffen}}(1990)}]{Steffen:1990abc}
\bibinfo{author}{\bibfnamefont{M.}~\bibnamefont{{Steffen}}},
  \bibinfo{journal}{Astronomy and Astrophysics} \textbf{\bibinfo{volume}{239}},
  \bibinfo{pages}{443} (\bibinfo{year}{1990}).

\bibitem[{\citenamefont{Galassi et~al.}(2019)}]{GSL:2019}
\bibinfo{author}{\bibfnamefont{M.}~\bibnamefont{Galassi}} \bibnamefont{et~al.}
  (\bibinfo{year}{2019}), \urlprefix\url{https://www.gnu.org/software/gsl/}.

\bibitem[{\citenamefont{Akaike}(1971)}]{Akaike:1973abc}
\bibinfo{author}{\bibfnamefont{H.}~\bibnamefont{Akaike}}, \bibinfo{journal}{in
  Petrov, B. N.; Csáki, F. (eds.), 2nd International Symposium on Information
  Theory, Tsahkadsor, Armenia, USSR, September 2-8} p. \bibinfo{pages}{pp.
  267–281.} (\bibinfo{year}{1971}).

\bibitem[{\citenamefont{Borsanyi et~al.}(2021)}]{Borsanyi:2020mff}
\bibinfo{author}{\bibfnamefont{S.}~\bibnamefont{Borsanyi}}
  \bibnamefont{et~al.}, \bibinfo{journal}{Nature}
  \textbf{\bibinfo{volume}{593}}, \bibinfo{pages}{51} (\bibinfo{year}{2021}),
  \eprint{2002.12347}.

\bibitem[{\citenamefont{Della~Morte et~al.}(2005)\citenamefont{Della~Morte,
  Hoffmann, Knechtli, Sommer, and Wolff}}]{DellaMorte:2005xgj}
\bibinfo{author}{\bibfnamefont{M.}~\bibnamefont{Della~Morte}},
  \bibinfo{author}{\bibfnamefont{R.}~\bibnamefont{Hoffmann}},
  \bibinfo{author}{\bibfnamefont{F.}~\bibnamefont{Knechtli}},
  \bibinfo{author}{\bibfnamefont{R.}~\bibnamefont{Sommer}}, \bibnamefont{and}
  \bibinfo{author}{\bibfnamefont{U.}~\bibnamefont{Wolff}},
  \bibinfo{journal}{JHEP} \textbf{\bibinfo{volume}{07}}, \bibinfo{pages}{007}
  (\bibinfo{year}{2005}), \eprint{hep-lat/0505026}.

\bibitem[{\citenamefont{Dalla~Brida et~al.}(2019)\citenamefont{Dalla~Brida,
  Korzec, Sint, and Vilaseca}}]{DallaBrida:2018tpn}
\bibinfo{author}{\bibfnamefont{M.}~\bibnamefont{Dalla~Brida}},
  \bibinfo{author}{\bibfnamefont{T.}~\bibnamefont{Korzec}},
  \bibinfo{author}{\bibfnamefont{S.}~\bibnamefont{Sint}}, \bibnamefont{and}
  \bibinfo{author}{\bibfnamefont{P.}~\bibnamefont{Vilaseca}},
  \bibinfo{journal}{Eur. Phys. J. C} \textbf{\bibinfo{volume}{79}},
  \bibinfo{pages}{23} (\bibinfo{year}{2019}), \eprint{1808.09236}.

\bibitem[{\citenamefont{Vuorinen}(2003)}]{Vuorinen:2002ue}
\bibinfo{author}{\bibfnamefont{A.}~\bibnamefont{Vuorinen}},
  \bibinfo{journal}{Phys. Rev. D} \textbf{\bibinfo{volume}{67}},
  \bibinfo{pages}{074032} (\bibinfo{year}{2003}), \eprint{hep-ph/0212283}.

\bibitem[{\citenamefont{T\"or\"ok et~al.}(2021)\citenamefont{T\"or\"ok, C\`e,
  Harris, Krasniqi, Meyer, and Toniato}}]{Torok:2021ujr}
\bibinfo{author}{\bibfnamefont{C.}~\bibnamefont{T\"or\"ok}},
  \bibinfo{author}{\bibfnamefont{M.}~\bibnamefont{C\`e}},
  \bibinfo{author}{\bibfnamefont{T.}~\bibnamefont{Harris}},
  \bibinfo{author}{\bibfnamefont{A.}~\bibnamefont{Krasniqi}},
  \bibinfo{author}{\bibfnamefont{H.~B.} \bibnamefont{Meyer}}, \bibnamefont{and}
  \bibinfo{author}{\bibfnamefont{A.}~\bibnamefont{Toniato}},
  \bibinfo{journal}{PoS} \textbf{\bibinfo{volume}{LATTICE2021}},
  \bibinfo{pages}{172} (\bibinfo{year}{2021}), \eprint{2112.00563}.

\bibitem[{\citenamefont{Meyer}(2018)}]{Meyer:2018xpt}
\bibinfo{author}{\bibfnamefont{H.~B.} \bibnamefont{Meyer}},
  \bibinfo{journal}{Eur. Phys. J. A} \textbf{\bibinfo{volume}{54}},
  \bibinfo{pages}{192} (\bibinfo{year}{2018}), \eprint{1807.00781}.

\bibitem[{\citenamefont{C\`e et~al.}(2021)\citenamefont{C\`e, Harris, Meyer,
  Toniato, and T\"or\"ok}}]{Ce:2021nvy}
\bibinfo{author}{\bibfnamefont{M.}~\bibnamefont{C\`e}},
  \bibinfo{author}{\bibfnamefont{T.}~\bibnamefont{Harris}},
  \bibinfo{author}{\bibfnamefont{H.~B.} \bibnamefont{Meyer}},
  \bibinfo{author}{\bibfnamefont{A.}~\bibnamefont{Toniato}}, \bibnamefont{and}
  \bibinfo{author}{\bibfnamefont{C.}~\bibnamefont{T\"or\"ok}}
  (\bibinfo{year}{2021}), \eprint{2112.00450}.

\bibitem[{\citenamefont{Fousse et~al.}(2007)\citenamefont{Fousse, Hanrot,
  Lef\`{e}vre, P\'{e}lissier, and Zimmermann}}]{10.1145/1236463.1236468}
\bibinfo{author}{\bibfnamefont{L.}~\bibnamefont{Fousse}},
  \bibinfo{author}{\bibfnamefont{G.}~\bibnamefont{Hanrot}},
  \bibinfo{author}{\bibfnamefont{V.}~\bibnamefont{Lef\`{e}vre}},
  \bibinfo{author}{\bibfnamefont{P.}~\bibnamefont{P\'{e}lissier}},
  \bibnamefont{and}
  \bibinfo{author}{\bibfnamefont{P.}~\bibnamefont{Zimmermann}},
  \bibinfo{journal}{ACM Trans. Math. Softw.} \textbf{\bibinfo{volume}{33}},
  \bibinfo{pages}{13–es} (\bibinfo{year}{2007}).

\bibitem[{\citenamefont{Granlund and the GMP~development team}(2020)}]{GMP:web}
\bibinfo{author}{\bibfnamefont{T.}~\bibnamefont{Granlund}} \bibnamefont{and}
  \bibinfo{author}{\bibnamefont{the GMP~development team}}
  (\bibinfo{year}{2020}), \urlprefix\url{https://gmplib.org/}.

\bibitem[{\citenamefont{Deur et~al.}(2016)\citenamefont{Deur, Brodsky, and
  de~Teramond}}]{Deur:2016tte}
\bibinfo{author}{\bibfnamefont{A.}~\bibnamefont{Deur}},
  \bibinfo{author}{\bibfnamefont{S.~J.} \bibnamefont{Brodsky}},
  \bibnamefont{and} \bibinfo{author}{\bibfnamefont{G.~F.}
  \bibnamefont{de~Teramond}}, \bibinfo{journal}{Nucl. Phys.}
  \textbf{\bibinfo{volume}{90}}, \bibinfo{pages}{1} (\bibinfo{year}{2016}),
  \eprint{1604.08082}.

\bibitem[{\citenamefont{Aoki et~al.}(2020)}]{Aoki:2019cca}
\bibinfo{author}{\bibfnamefont{S.}~\bibnamefont{Aoki}} \bibnamefont{et~al.}
  (\bibinfo{collaboration}{Flavour Lattice Averaging Group}),
  \bibinfo{journal}{Eur. Phys. J. C} \textbf{\bibinfo{volume}{80}},
  \bibinfo{pages}{113} (\bibinfo{year}{2020}), \eprint{1902.08191}.

\end{thebibliography}

\end{document}